\documentclass[11pt]{article}
\usepackage{authblk}
\usepackage{setspace}

\usepackage{hyperref}
\hypersetup{
colorlinks=true,
urlcolor=blue,
citecolor=blue}
\usepackage[all]{xy,xypic}
\usepackage{amsfonts,amssymb,amsmath,amsgen,amsopn,amsbsy,theorem,graphicx,epsfig}
\usepackage{eufrak,amscd,bezier,latexsym,mathrsfs,eurosym,enumerate}
\usepackage[utf8]{inputenc}
\usepackage[english]{babel}
\usepackage{cleveref,multirow}
\usepackage[dvipsnames]{xcolor}
\usepackage[pagewise]{lineno}
\usepackage{float}
\usepackage[caption = false]{subfig}

\usepackage[
    left=25 mm,
    right=25 mm,
    top=25 mm,
    bottom=25 mm
]{geometry}

\begin{document}
\onehalfspacing

\title{Mechatronic Design, Dynamic Modeling, and Real-Time Control of a Movable Scaffold}

\author[1]{M. Özgün GÜLEÇ \thanks{Current address: Dept. of Mechatronics Engineering, İzmir University of Economics, İzmir, Turkey, e-mail: \texttt{ozgun.gulec@izmirekonomi.edu.tr}}}
\author[2]{Koray K. ŞAFAK}
\affil[1,2]{Dept. of Mechanical Engineering, Yeditepe University, Istanbul, Turkey}
\affil[2]{e-mail: \texttt{safak@yeditepe.edu.tr}}

\maketitle

\begin{abstract} This study presents mechatronic design, dynamic modeling, simulations and real-time control experiments of a new movable scaffolding system. The proposed system consists of a 3 degrees-of-freedom movable platform, which can be positioned on the outer surface of buildings. The platform is supported and driven by cords that are wound on pulleys and coupled to servo controlled dc-motors located at four corners of the building surface. A mathematical model considering the actuator dynamics for this cable-driven mechanism is obtained and its simulation results are presented. Design, manufacture and real-time control tests of a prototype has been done. Both numerical simulations and experiments provide good positioning performance of the proposed cable-driven mechanism.
\end{abstract}


\section{Introduction}
\label{Int}
Scaffolding has been widely used in construction, repair or cleaning operations of buildings and large structures. Most of the scaffolding is stationary structures that are being built on the outer surface of the building that is to be worked on. Movable scaffolding systems can replace the stationary ones by offering a flexible and more convenient solution for the same purpose. The movable scaffolding is a relatively new approach and there are not so many solutions offered to achieve movable scaffolding operations.
The proposed movable scaffolding system is in the form of a cable-supported robot. Cable-supported robots have been widely studied over the last few decades \cite{RN123, RN124, RN125, RN122}; however these robots have not been utilized for scaffolding operations on buildings. Easy assembly, modular structure and dexterity of cable-suspended robots make them attractive for a number of different uses. One important issue that needs to be addressed is to keep track of the cable tensions, which has to be always positive. A number of cable-suspended planar robots considering various aspects of design, control, and purpose of use have been reported. 

Cable-driven parallel robots have emerged as a promising field of research and innovation in robotics. These robots, which utilize cables as their primary means of actuation, offer unique advantages in terms of lightweight construction, low power consumption, and high flexibility. In recent years, a significant body of research has been dedicated to the development, control, and analysis of cable-driven parallel robots, particularly planar ones \cite{RN121, RN109, RN129}. The concept of cable-driven parallel robots has evolved significantly since its inception. Early work by Albus et al. \cite{RN124} introduced the concept of CDPRs as a means to achieve large workspace areas with minimal structural weight. 

Cable-driven parallel robots, often referred to as cable-driven manipulators, are characterized by their ability to manipulate objects through cable tension. These robots are extensively used in various applications, including cleaning high-rise building exteriors \cite{RN111}, radio telescope feed platform control \cite{RN117}, and high-speed manipulation \cite{RN118} tasks. One of the key advantages of cable-driven robots is their inherent redundancy, which allows for versatility in motion and enhanced workspace capabilities.

Various types of feedback control methods for the cable-driven platforms have been studied. In \cite{RN116}, an adaptive robust control algorithm for fully-constrained cable-driven parallel robots is presented. This algorithm ensures cable tension, exhibits computational efficiency, and maintains bounded tracking error. Experimental verification highlights its practical effectiveness. Another research investigates the stiffness of cable-based robots \cite{RN120}, focusing on antagonistic forces and their role in design and control. It introduces conditions for achieving system stability, especially critical for space applications. Dynamic modeling and simulation of these systems are needed for precise control applications. In \cite{RN114}, the dynamic analysis and control of cable-driven redundant parallel manipulators with flexible cables is studied. It models cables as axial linear springs, introduces internal forces, and develops a control algorithm for cable-driven robots. Position control control of a fully constrained CDPR with unkown or partially known dynamics is also studied  \cite{RN108}. This research introduces an intelligent proportional-integral-derivative (iPID) controller for position control of planar cable-driven robots. It adapts to system dynamics with partially known information and enhances control performance. Reinforcement learning methods have also been utilized in control of planar CDPR's \cite{RN110}. In \cite{RN112} adaptive cable-driven 3DOF parallel robot with cable wrapping phenomenon has been studied. The cable wrapping phenomenon is harnessed to enhance workspace and dexterity.

Cable suspended planar robots with redundant cables have also been proposed \cite{RN125}. Such systems with redundant cables require a control system, which will keep the cable tensions positive. An optimization approach for the design of planar redundant cable-suspended robots, while considering the minimum time trajectory tracking criterion, have been presented \cite{RN126}. Design and control of two types of planar cable driven robots for upper limb neuro-rehabilitation of stroke patients have been proposed \cite{RN127}. Dynamics and control of a cable-suspended parallel robot with six degrees of freedom, intended for positioning of large spherical radio telescopes have been presented \cite{RN113}. In that study, the effect of gravity on the shape of the cables has also been considered. Analysis and control of the cable-supported six degrees of freedom parallel robot considering the effect of the actuator dynamics have also been studied \cite{RN128}. Skycam cable-suspended mechanism is used commercially in the stadiums of several big sports clubs \cite{RN123}. This mechanism consists of a camera holder platform, which is supported by cables attached at the four high fixed points of the stadium and driven by motors.

Currently, the scaffolding systems used on buildings are mostly static structures composed of number of trusses and beams. These trusses and beams form a structure which is actually very hard to assemble and disassemble and these types of structures are usually unsuccessful at reaching every point on the external surface. In the present study, a 3-DOF cable-supported planar scaffold system, which is intended to operate on the outside surfaces of buildings and large structures, is presented. The system consists of a movable platform that is driven by coupled electric motors and pulleys by winding and unwinding the suspension cables. Remaining sections of the paper can be outlined as follows. Section 2 describes the mechatronic design details and implementation of a prototype. Kinematic and dynamic modeling of the system and its simulation results are presented in Section 3. Testing of the prototype and experimental results are discussed in Section 4. Conclusions are drawn in the last section.

\section{Mechatronic Design of the System} 
\begin{figure}[h!]
\centering
\includegraphics[width=8.0cm]{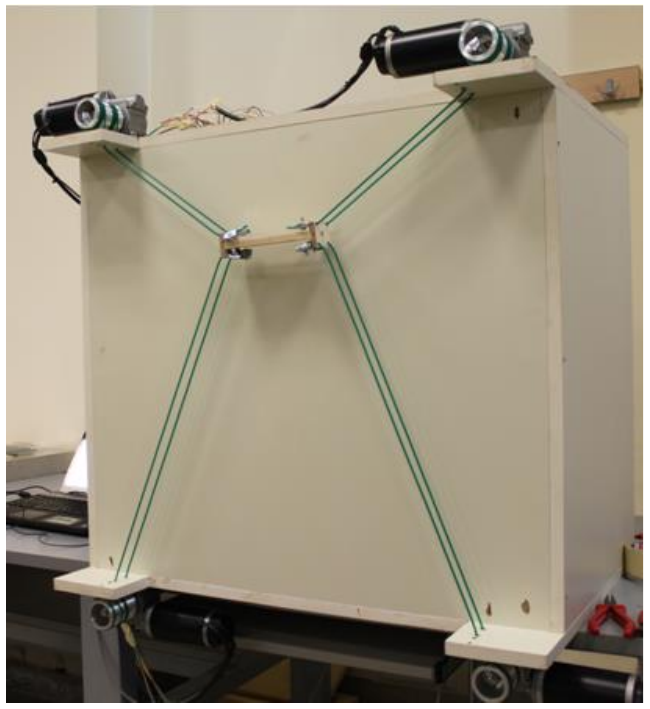}
\caption{Prototype of the system}
\label{fig1}
\end{figure}

\begin{figure}[h!]
\centering
\includegraphics[width=9.0cm]{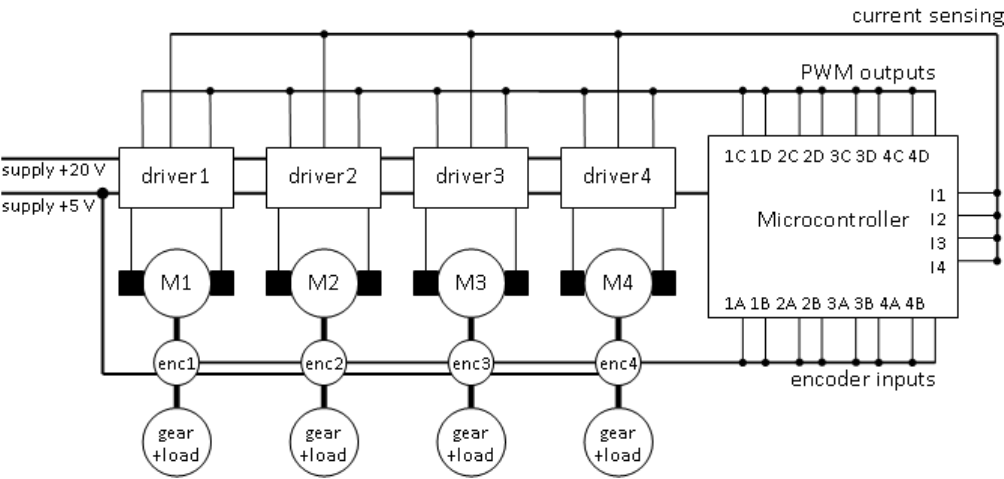}
\caption{Schematic diagram of the electrical components}
\label{fig2}
\end{figure}

The movable scaffold system consists of a rectangular plate supported by belts at the corners. These belts are coiled and released by pulleys that are driven by DC motors. The motor-pulley arrangements are fixed at four locations on a wooden stand, which represents the operating environment of the movable scaffold (see Figure~\ref{fig1}).
A microcontroller board (Arduino Mega 2560) based on ATmega2560 (of Atmel Co.) is used as the main control unit. The unit has 54 digital input/output pins (of which 14 can be used as PWM outputs), 16 analog inputs, 4 UARTs (hardware serial ports), a 16 MHz crystal oscillator, a USB connection, a power jack, an ICSP header, and a reset button. Usage of the microcontroller for the planar movable scaffold system serves for 3 different purposes. The main purpose is to generate the PWM signal for the motor driver circuit. In order to operate the motor in both clockwise and counterclockwise directions, every DC motor needs 2 PWM signals, so 8 PWM signals are taken from the microcontroller. The second purpose is for the encoder readings. DC motors have encoders to provide position feedback. For accurate positioning, the microcontroller needs to collect encoder data and use this data for correcting the PWM signals' magnitude that controls the motor speed and direction. The encoder pins are connected to the interrupt and digital pins of the microcontroller. Lastly, microcontroller analog inputs are utilized to get current measurements on DC motors. Hence, the current characteristics of the DC motors can be monitored while they operate.
Each DC motor is driven by an L298 full-bridge circuit (from STMicroelectronics Co.) with outputs in parallel configuration. This circuit consists of a pair of motor drivers with a continuous current rating of 2 A. When paralleled, the driver circuit can provide continuous currents of up to 4 A. The DC motor used in planar movable scaffold provides maximum 2Nm torque with 193 rpm speed, and consumes 3.5A at nominal conditions. Incremental encoders sense the rotation of the DC motors and convert the rotation to the digital pulses including the rotation direction information of the motor. The encoders provide position feedback with a resolution of 400 pulses per revolution of the motor shaft. Motors are positioned and mounted to the stand such that coordinates of the cords are compatible with the coordinates defined in the control algorithm. The cables are cut by considering the platform dimensions and then eight cables with equal lengths are obtained. These cables provide the transmission of the movements and the power from pulleys to the platform. Every pulley coils two cables and, other sides of the cables are connected to one corner of the platform.
A sensing resistor soldered between the ground and the motor driver is used for measuring and limiting the actual current drawn by motors. There are two capacitors used in this circuit one of them is soldered between ground and voltage supply pin and another capacitor is soldered between ground and logic voltage level pin. Every L298 takes two PWM signals from the microcontroller and generates high-current driving signals for the DC motors. The electrical voltage requirements of 20 V for the supply and 5 V for the enable signals of L298 have been provided by two DC power supplies and by the microcontroller, respectively. The microcontroller also provides 5 DC volts to every encoder and reads the encoders’ data from interrupt and digital pins. The current sensing wires are connected to the analog input pins to monitor the currents drawn by the motors. A schematic diagram of the electrical components is shown in Figure~\ref{fig2}.

\subsection{Real-time Control Scheme} 
\begin{figure}[h!]
\centering
\includegraphics[width=9.0cm]{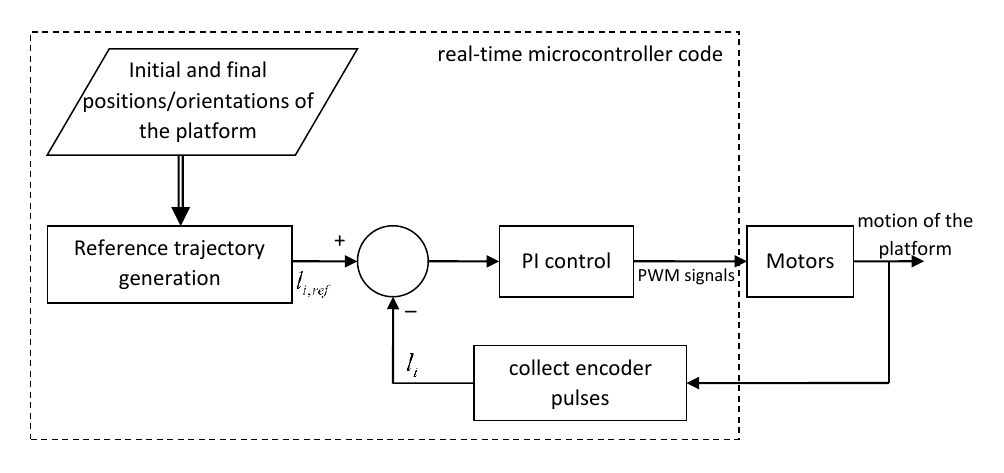}
\caption{Blockdiagram of the real-time microcontroller code}
\label{fig3}
\end{figure}

A proportional-integral (PI) control scheme is utilized in both experimental and simulation studies. For the experimental part, the mathematical formulas for the cords are redesigned and a reference generation function is created in the microcontroller software. After the initial and final positions of the platform are selected, the microcontroller code executes and generates reference lengths for all four cords, throughout the duration of travel. After generating one reference, software reads information from all four encoders and converts them to the actual lengths. PI control tries to minimize the difference between actual and reference lengths of the cords. Depending on the absolute values of differences and the integrals of the differences, the quantities are multiplied with the Kp and Ki parameters to generate the necessary PWM signals. These Kp and Ki parameters are selected experimentally as 0.9 and 0.01, respectively based on some manual tuning. The software continues to generate reference until starting time reaches the calculated travel time. Blockdiagram of the real-time microcontroller code is shown in Figure~\ref{fig3}.

\section{Dynamic Modeling and Simulations} 
\begin{figure}[h]
\centering
\includegraphics[width=8.0cm]{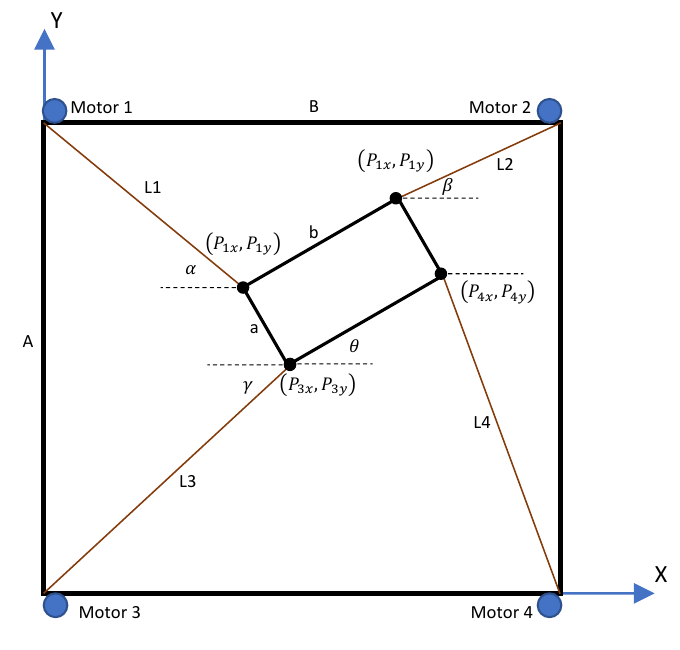}
\caption{Schematic diagram and parameters of the movable scaffold}
\label{fig4}
\end{figure}

A schematic diagram of the movable scaffold system is shown in Figure~\ref{fig4}. The platform position in fixed XY coordinate system and orientation are defined by Px, Py, and, ${\theta}$, respectively. Lengths of cords are defined by L1, L2, L3 and L4. Positions of the corners in XY coordinates are $P_{ix}$, $P_{iy}$, where $i$ indicates the corner number. Height and width parameters for the stand and for the platform are denoted by A, B, a, b, respectively. Angle between cords 1, 2, 3, and X{}-axis are defined by ${\alpha}$, ${\beta}$, and ${\gamma}$, respectively.
The mechanism position and orientation of the platform (i.e., Px, Py, and, ${\theta}$) are fixed, whenever three of the four supporting cords (for example L1, L2, and L3) have prescribed lengths. This condition is verified by the Grübler's equation, which is commonly used for kinematic analysis of mechanisms. For this analysis, the cords are assumed to be rigid links with prescribed lengths. Secondly, the attachment points of the cords to both the movable platform and the stand are assumed to be type 1 joints. Hence, they allow rotation of the cords about these attachment points. Grübler's equation is stated as:  $f=3n-2j_1-j_2-3$, where $f$ is the number of net degrees of freedom, $n$ is number of links, $j_1$ is the number of pin joints, and $j_2$ is the number of cylindrical joints. It is shown that the proposed system has only one configuration for a given set of cord lengths ($f=3\ast 5-2\ast 6-3=0$), hence it has zero degrees of freedom and locked.
\subsection{Kinematic Equations} Equations (1-3) are generated according to the instantaneous position of the movable platform, as seen in Figure~\ref{fig4}. The parameter  $P_{1x}$ \ stands for the x coordinate of the first edge of the movable platform. The edges are numbered according to the cords that are attached to them.
\begin{equation}
P_{1x}=P_x-\frac b 2\cos \left(\theta \right)-\frac a 2\sin \left(\theta \right)
\end{equation}
\begin{equation}
\frac{dP_{1x}}{\mathit{dt}}=\frac{dP_x}{\mathit{dt}}+\frac b 2\sin \left(\theta \right)\frac{\mathit{d\theta
}}{\mathit{dt}}-\frac a 2\cos \left(\theta \right)\frac{\mathit{d\theta }}{\mathit{dt}}
\end{equation}
\begin{equation}
\frac{d^2P_{1x}}{dt^2}=\frac{d^2P_x}{dt^2}+\frac b 2\cos \left(\theta \right)\left(\frac{\mathit{d\theta
}}{\mathit{dt}}\right)^2+\frac b 2\sin \left(\theta \right)\frac{d^2\theta }{dt^2}+\frac a 2\sin \left(\theta
\right)\left(\frac{\mathit{d\theta }}{\mathit{dt}}\right)^2-\frac a 2\cos \left(\theta \right)\frac{d^2\theta
}{dt^2}
\end{equation}
Equation (1) is generated by using geometric relations, whereas equations (2) and (3) are derived by taking the first and second derivatives of equation (1), respectively.
\begin{equation}
P_{1y}=P_y+\frac a 2\cos \left(\theta \right)-\frac b 2\sin \left(\theta \right)
\end{equation}
\begin{equation}
\frac{dP_{1y}}{\mathit{dt}}=\frac{dP_y}{\mathit{dt}}-\frac a 2\sin \left(\theta \right)\frac{\mathit{d\theta
}}{\mathit{dt}}-\frac b 2\cos \left(\theta \right)\frac{\mathit{d\theta }}{\mathit{dt}}
\end{equation}
\begin{equation}
\frac{d^2P_{1y}}{dt^2}=\frac{d^2P_y}{dt^2}-\frac a 2\cos \left(\theta \right)\left(\frac{\mathit{d\theta
}}{\mathit{dt}}\right)^2-\frac a 2\sin \left(\theta \right)\frac{d^2\theta }{dt^2}+\frac b 2\sin \left(\theta
\right)\left(\frac{\mathit{d\theta }}{\mathit{dt}}\right)^2-\frac b 2\cos \left(\theta \right)\frac{d^2\theta
}{dt^2}
\end{equation}
For the y coordinate of position, velocity and acceleration of the first edge of the movable platform, equations (4-6) are generated. Angle between the first cord and the edge of the platform, ${\alpha}$ is expressed by equation (7). Equations (8-10) define instantaneous position, velocity and acceleration for cord number 1. Equation (8) is derived from geometry, whereas equations (9) and (10) are derived by taking the first and second derivatives of equation (8), respectively.
\begin{equation}
\alpha =\tan ^{-1}\left(\frac{A-P_{1y}}{P_{1x}}\right)
\end{equation}
\begin{equation}
l_1=\sqrt{P_{1x}^2+(A-P_{1y})^2}
\end{equation}
\begin{equation}
\frac{dl_1}{\mathit{dt}}=\frac{P_{1x}\frac{dP_{1x}}{\mathit{dt}}-(A-P_{1y})\frac{dP_{1y}}{\mathit{dt}}}{l_1}
\end{equation}
\begin{equation}
\frac{d^2l_1}{dt^2}=\frac{\left(\frac{dP_{1x}}{\mathit{dt}}\right)^2+P_{1x}\frac{d^2P_{1x}}{dt^2}+\left(\frac{dP_{1y}}{\mathit{dt}}\right)^2-\left(A-P_{1y}\right)\frac{d^2P_{1y}}{dt^2}-\left(\frac{dl_1}{\mathit{dt}}\right)^2}{l_1}
\end{equation}
Equation (11) relates the angular velocity of motor 1 output shaft to velocity of cord 1, whereas equation (12) relates the angular acceleration of motor 1 output shaft to acceleration of cord 1. In equations (11) and (12), r indicates the pulley radius.
\begin{equation}
\omega _1=\frac{-dq_1}{\mathit{dt}}=\frac{-\frac{dl_1}{\mathit{dt}}} r
\end{equation}
\begin{equation}
\frac{d\omega _1}{\mathit{dt}}=\frac{-d^2q_1}{dt^2}=\frac{-\frac{d^2l_1}{dt^2}} r
\end{equation}
Equations (13) and (14) define the overall dynamics of the movable platform. The second derivative of $P_x$ in equation (13) represents the acceleration of the platform in x direction, whereas  $P_y$  $P_y$ the second derivative of  $P_y$\ in equation (14) represents the acceleration of the platform in y direction.
\begin{equation}
-T_1\cos \left(\alpha \right)+T_2\cos \left(\beta \right)-T_3\cos \left(\gamma \right)-m\frac{d^2P_x}{dt^2}=0
\end{equation}
\begin{equation}
T_1\sin \left(\alpha \right)+T_2\sin \left(\beta \right)-T_3\sin \left(\gamma
\right)-\mathit{mg}-m\frac{d^2P_y}{dt^2}=0
\end{equation}
Equation (15) defines the sum of the moments caused by tensions  $T_1$,  $T_2$ \ and  $T_3$. In equation (15),  $I_p$ \ represents the inertia of the movable platform, and  $\frac{d^2\theta }{dt^2}$ \ represents the angular acceleration of the movable platform.
\begin{equation}
\frac{T_1a\cos \left(\alpha +\theta \right)} 2-\frac{T_1b\sin \left(\alpha +\theta \right)} 2-\frac{T_2a\cos \left(\beta
-\theta \right)} 2+\frac{T_2b\sin \left(\beta -\theta \right)} 2-\frac{T_3a\cos \left(\gamma -\theta \right)}
2+\frac{T_3b\sin \left(\gamma -\theta \right)} 2-I_p\frac{d^2\theta }{dt^2}=0
\end{equation}
While the platform is in motion, only three of the four motors provide actuation torques. As stated earlier, position and orientation of the platform can be defined by prescribing lengths of three supporting cords at any time. In other words, the fourth cable will be redundant. Whenever the platform is within the right half of the stand, the cord at the upper right corner does not carry tension. Similarly, whenever the platform stays within the left half of the structure, the cord at the upper left corner has no tension. Hence, whenever a cord has no tension the other three
cords are loaded. So, the system is driven by three of the four motors, depending on the location of the platform. The fourth motor will only synchronize its cord position, without carrying a significant load.
Required motor torques  $\tau _1$,  $\tau _2$, $\tau _3$ are modeled by equations (16-18). In these equations,  $\tau_0$ \ represents the dry friction torque,  $T_1,T_2,T_3$ \ are the tensions in the cords, $I$ is the pulley inertia, and c is the viscous damping constant that is calculated experimentally. The variables  $q_1,q_2,q_3$ \ are the angular positions of the output shafts. In the equations, first and second derivatives of q represent the angular velocity and acceleration of that motor shaft, respectively.
\begin{equation}
\tau _1=\mathit{sgn}\left(\frac{dq_1}{\mathit{dt}}\right)\tau _0+c\frac{dq_1}{\mathit{dt}}+T_1r-\left(\frac I r\right) \frac{d^2q_1}{dt^2}
\end{equation}
\begin{equation}
\tau _2=\mathit{sgn}\left(\frac{dq_2}{\mathit{dt}}\right)\tau _0+c\frac{dq_2}{\mathit{dt}}+T_2r-\left(\frac I r\right) \frac{d^2q_2}{dt^2}
\end{equation}
\begin{equation}
\tau _3=\mathit{sgn}\left(\frac{dq_3}{\mathit{dt}}\right)\tau _0+c\frac{dq_3}{\mathit{dt}}+T_3r-\left(\frac I r\right) \frac{d^2q_3}{dt^2}
\end{equation}
\subsection{Simulation Results} 
\begin{figure}[h!]
\centering
\includegraphics[width=9.0cm]{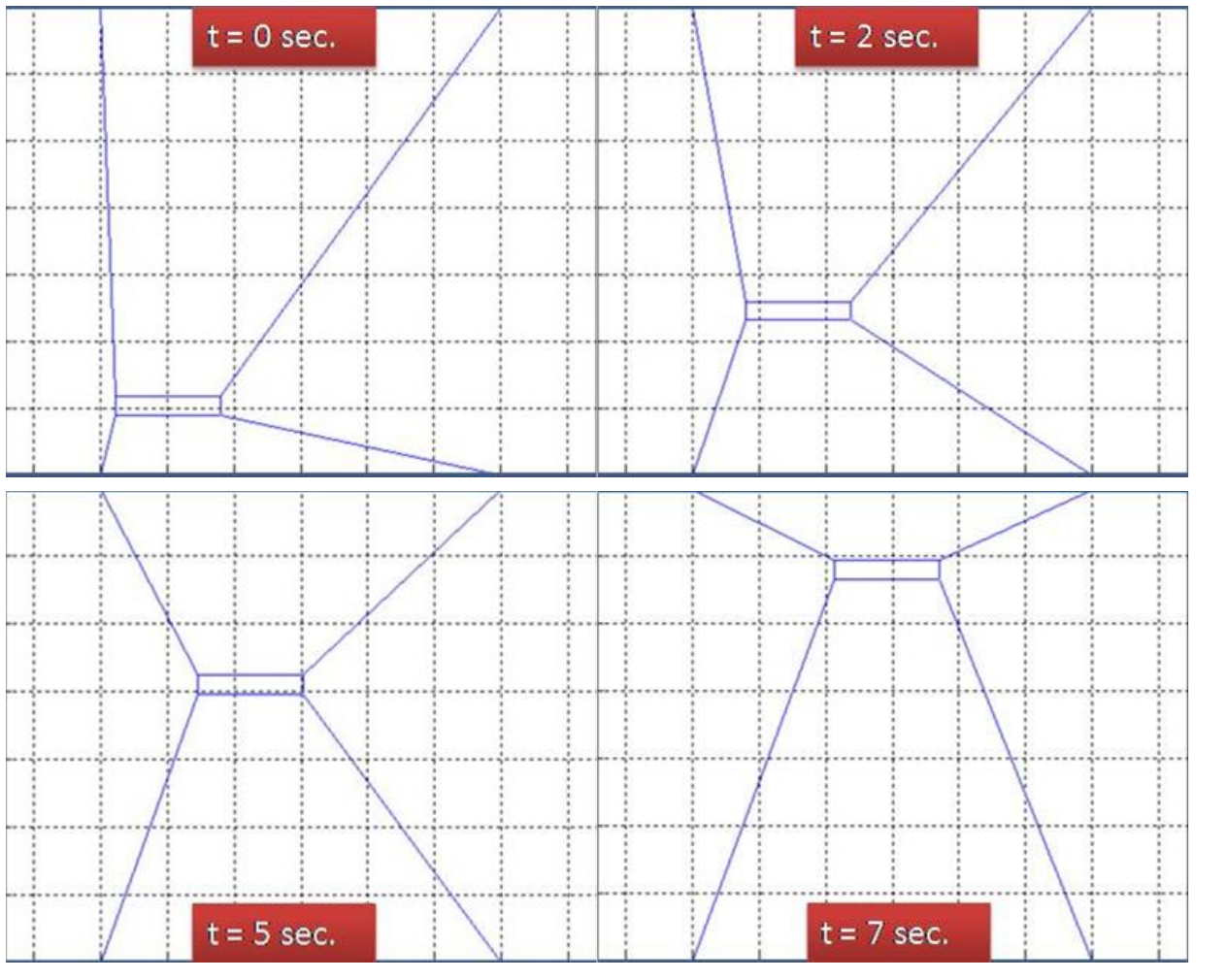}
\caption{Simulation results}
\label{fig5}
\end{figure}

\begin{figure}[h!]
\centering
\subfloat[]{\includegraphics[width = 5.0cm]{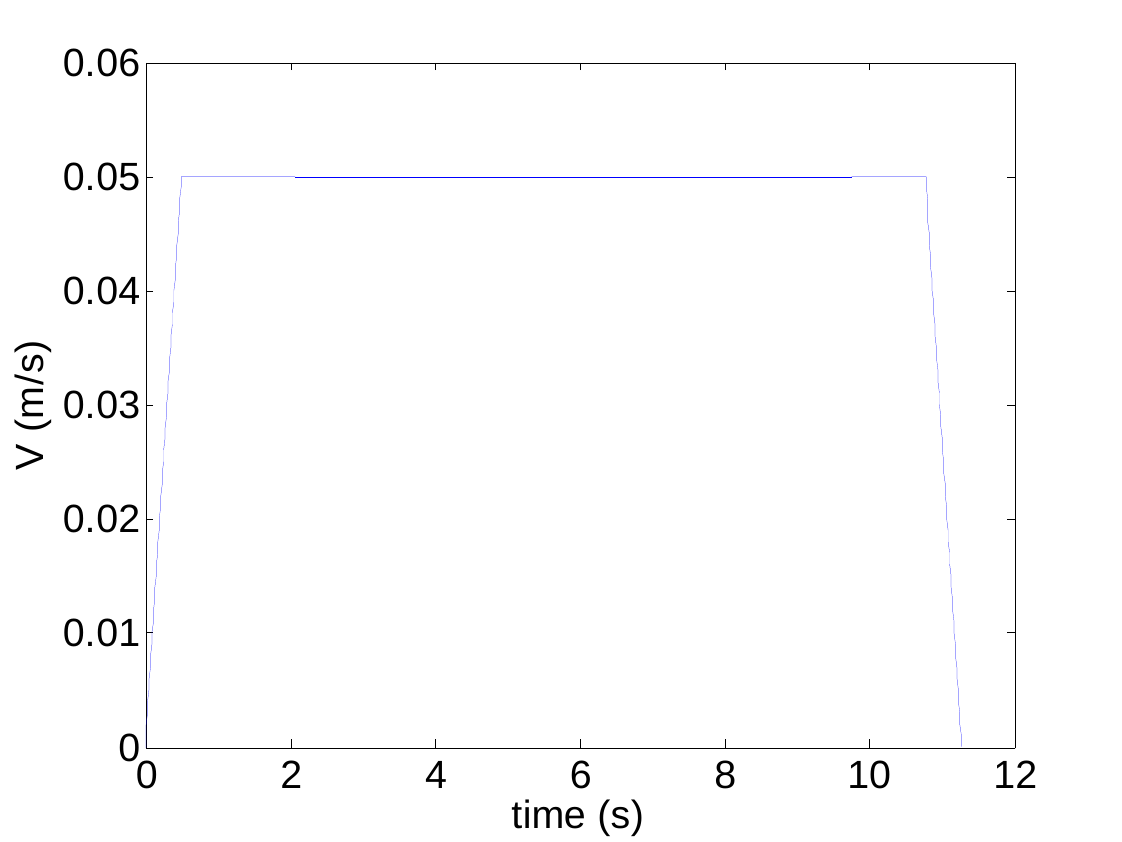}} 
\subfloat[]{\includegraphics[width = 5.0cm]{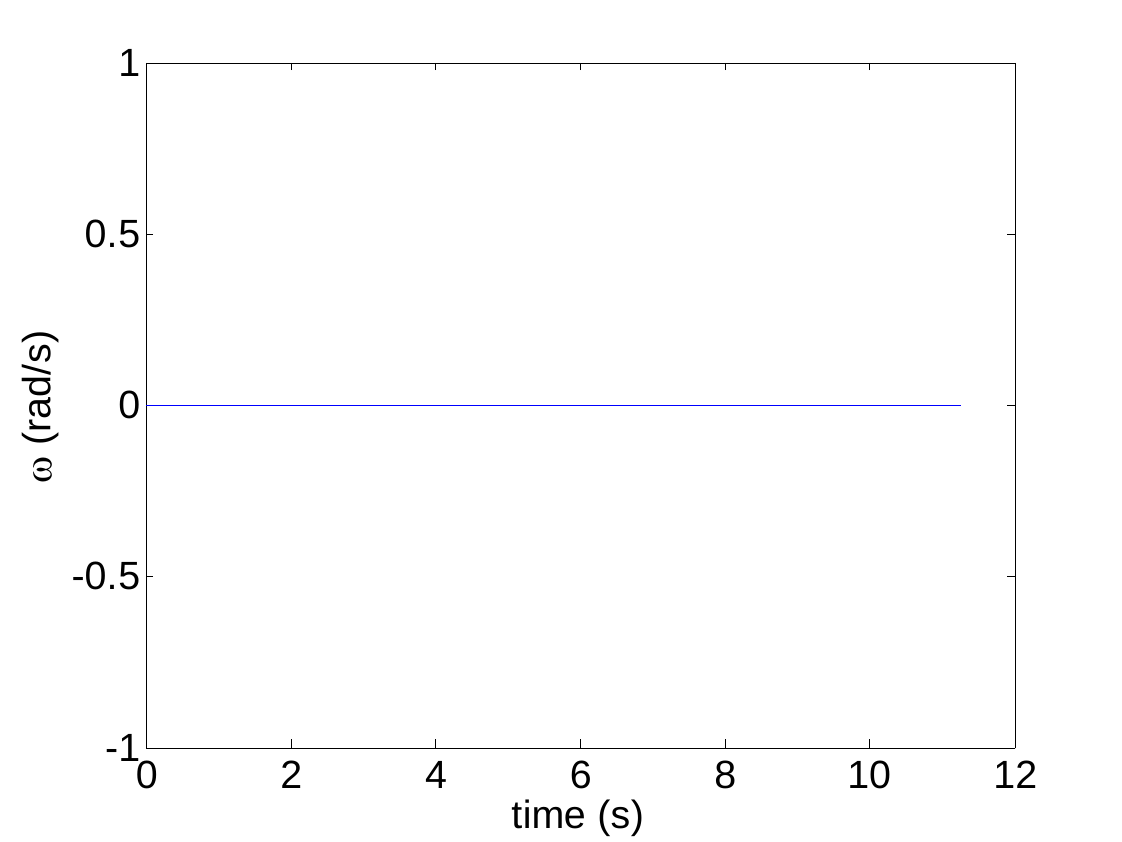}}\\
\subfloat[]{\includegraphics[width = 5.0cm]{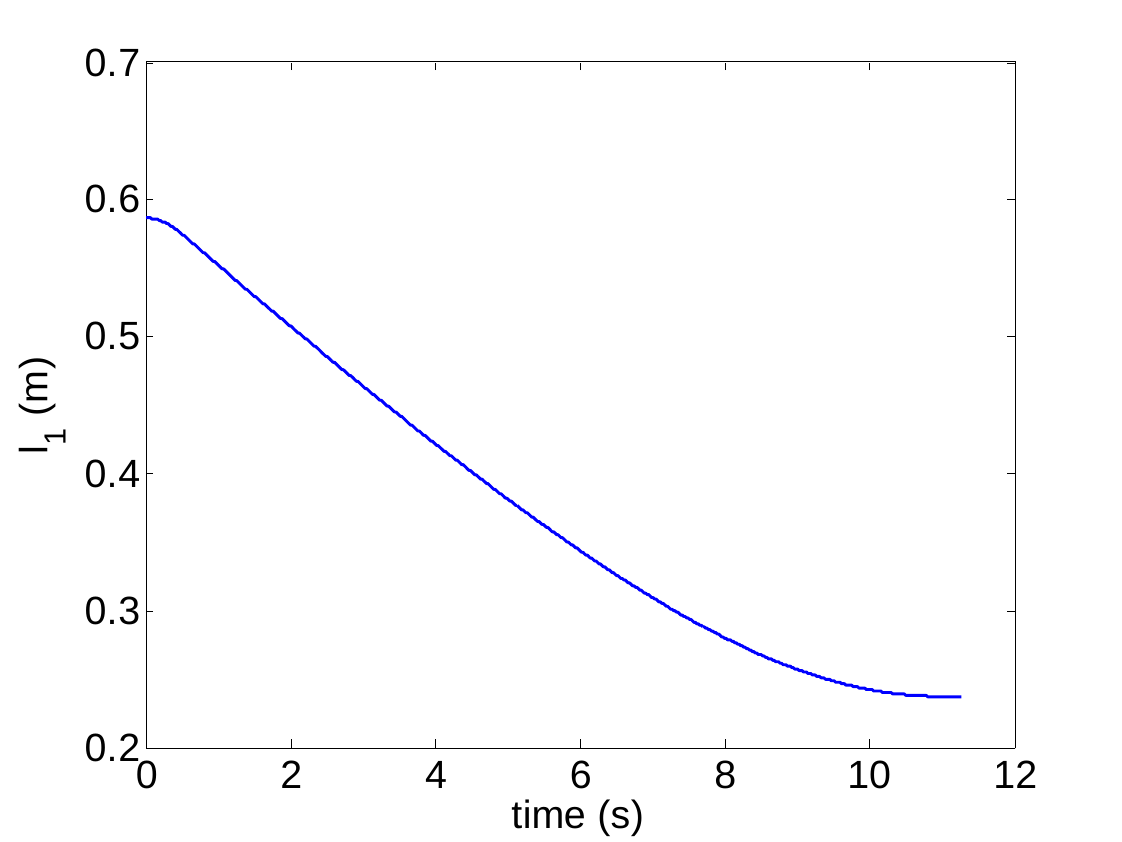}}
\subfloat[]{\includegraphics[width = 5.0cm]{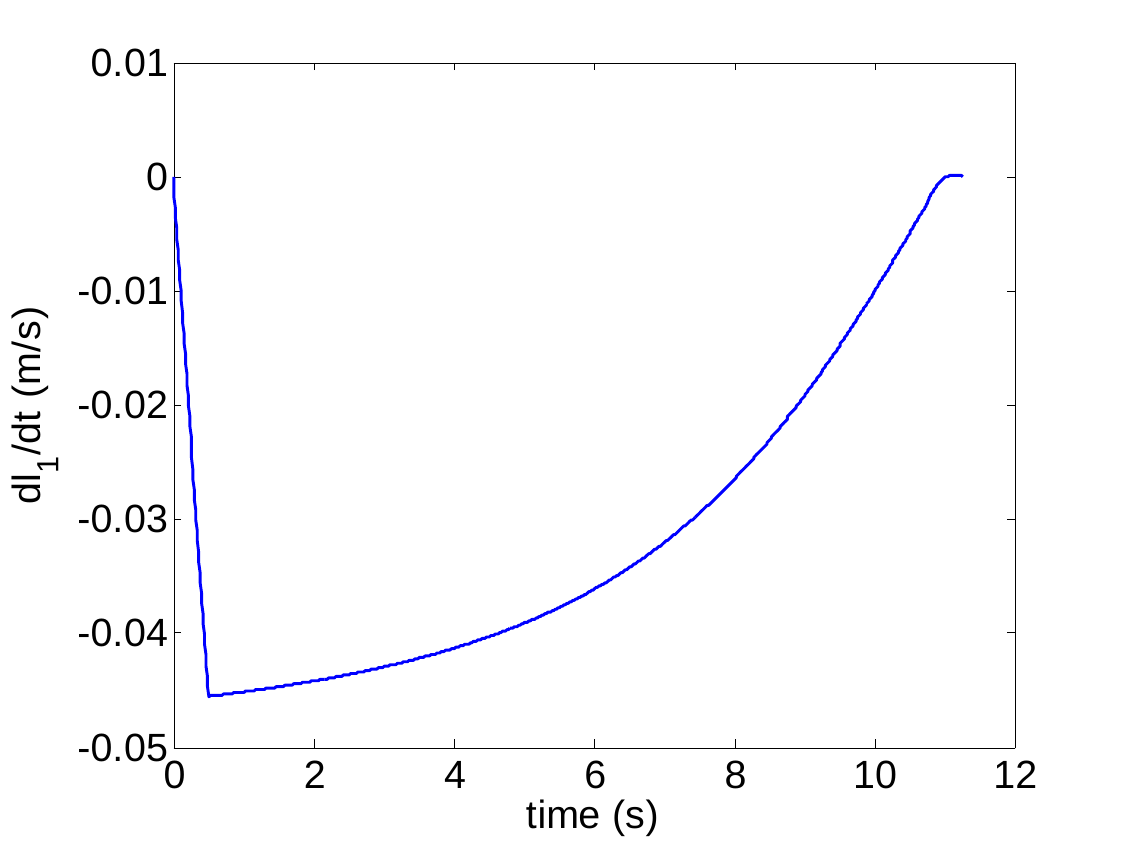}} 
\caption{Generated reference motion trajectory (a) translational and, (b) angular velocity reference for the platform, (c) reference length, and (d) reference velocity profiles for cord 1.}
\label{fig6}

\end{figure}

In this part, numerical simulation results for the movable scaffold system are presented. A reference trajectory for the platform position and corresponding trajectory for the cord lengths is generated. PI controllers are used to maintain the actual cord positions at the generated reference positions. An example simulation whose initial point (x, y) is (10, 10) cm and final point is (30, 60) cm with no angular motion (${\theta}$ = 0 rad) is run with a platform velocity of 0.05 m/sec and platform acceleration of 0.1 m/sec2. Figure~\ref{fig5} shows the movable platform at four instants during this simulation. It is observed that the platform moves diagonally while moving from its initial to final positions. During the simulation, the platform remains horizontal as it is commanded by the simulation program ($\theta $ = 0
rad).
Figure~\ref{fig6}a and Figure~\ref{fig6}b show the translational and angular velocity profiles for the platform, respectively. The platform is commanded a trapezoidal profile for its translational velocity. The acceleration and deceleration periods have the same rate. Since there is no commanded angle of the platform, there is no angular velocity indicated and this graph has a constant value of zero (Figure~\ref{fig6}b). Figure~\ref{fig6}c shows the cord length for motor 1 carrying the platform at one edge. In this figure, the cord length starts at around 0.6 m and shrinks down to about 0.25 m. The velocity profile during the motion is not constant and decreases with time (Figure~\ref{fig6}d). Parameters of the system are given in Table~\ref{tab1}.

\begin{figure}[h!]
\centering
\subfloat[]{\includegraphics[width = 6.0cm]{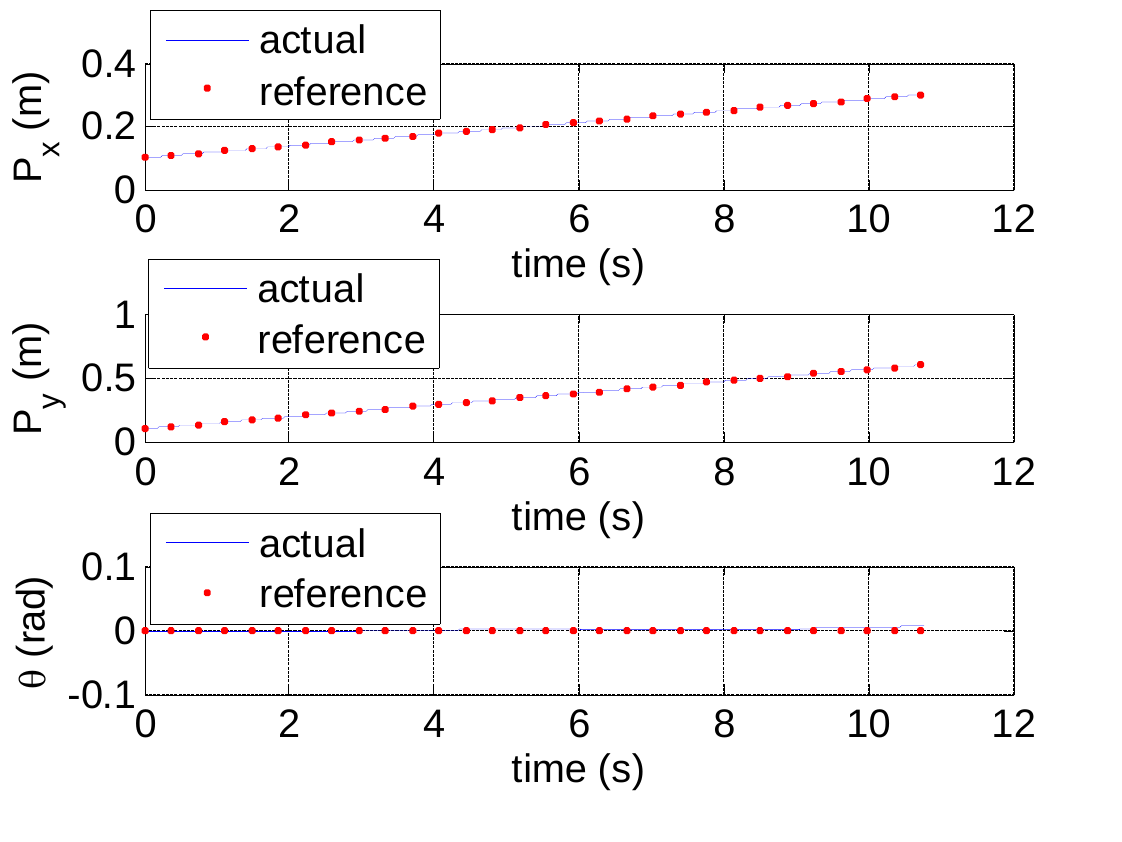}} 
\subfloat[]{\includegraphics[width = 6.0cm]{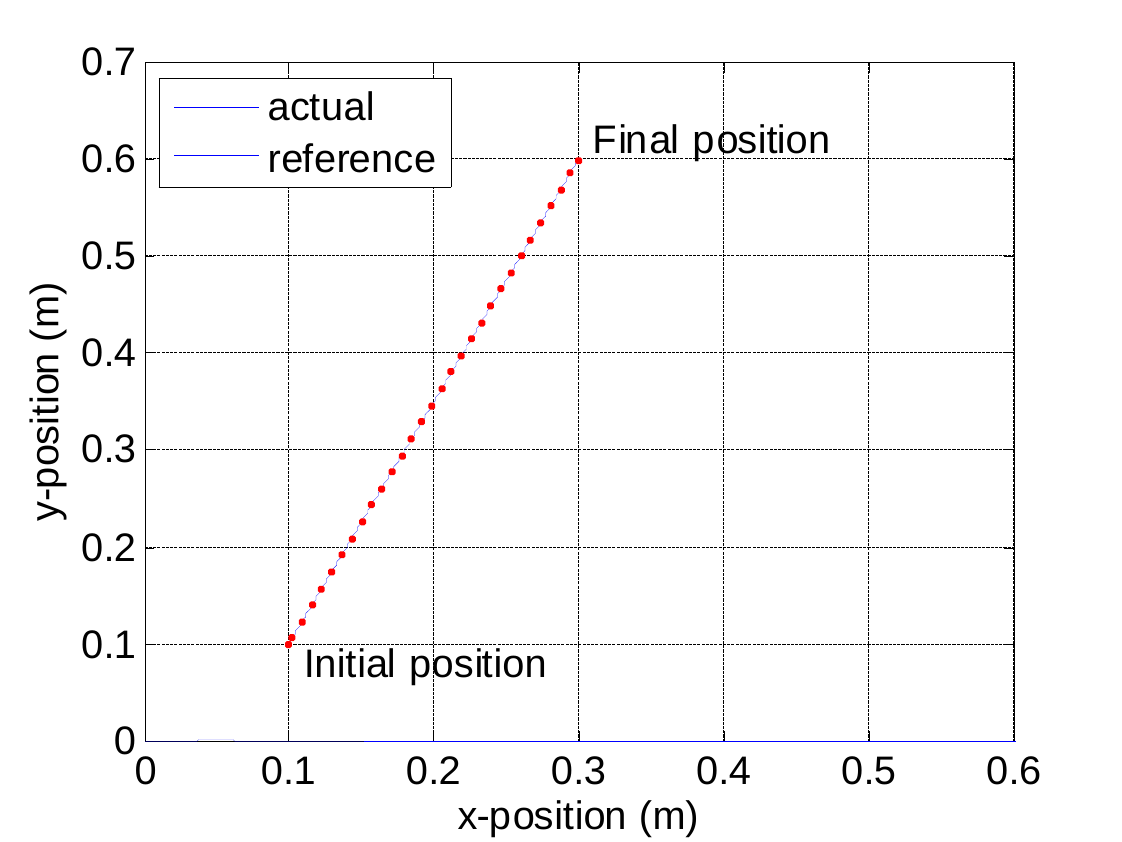}}\\
\caption{(a) Actual and reference positions of the platform, (b) platform trajectory}
\label{fig7}

\end{figure}

\begin{figure}[h!]
\centering
\subfloat[]{\includegraphics[width = 5.0cm]{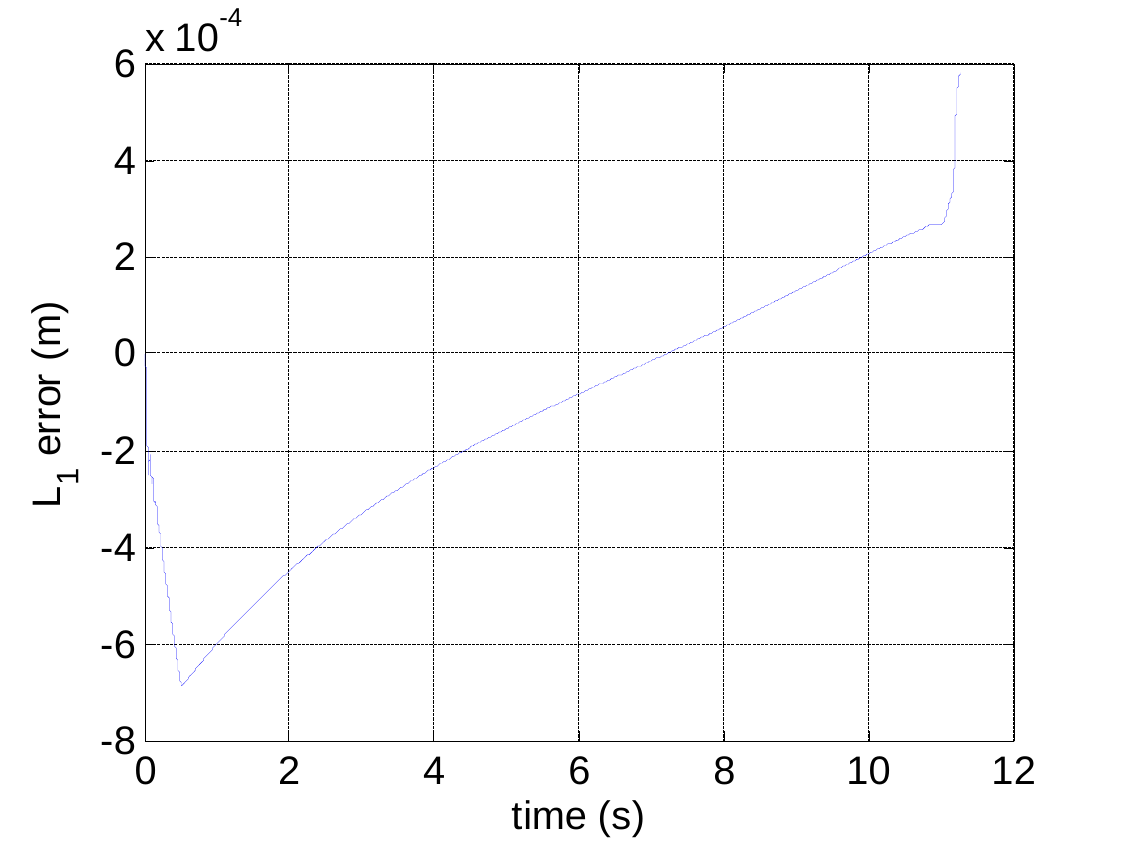}} 
\subfloat[]{\includegraphics[width = 5.0cm]{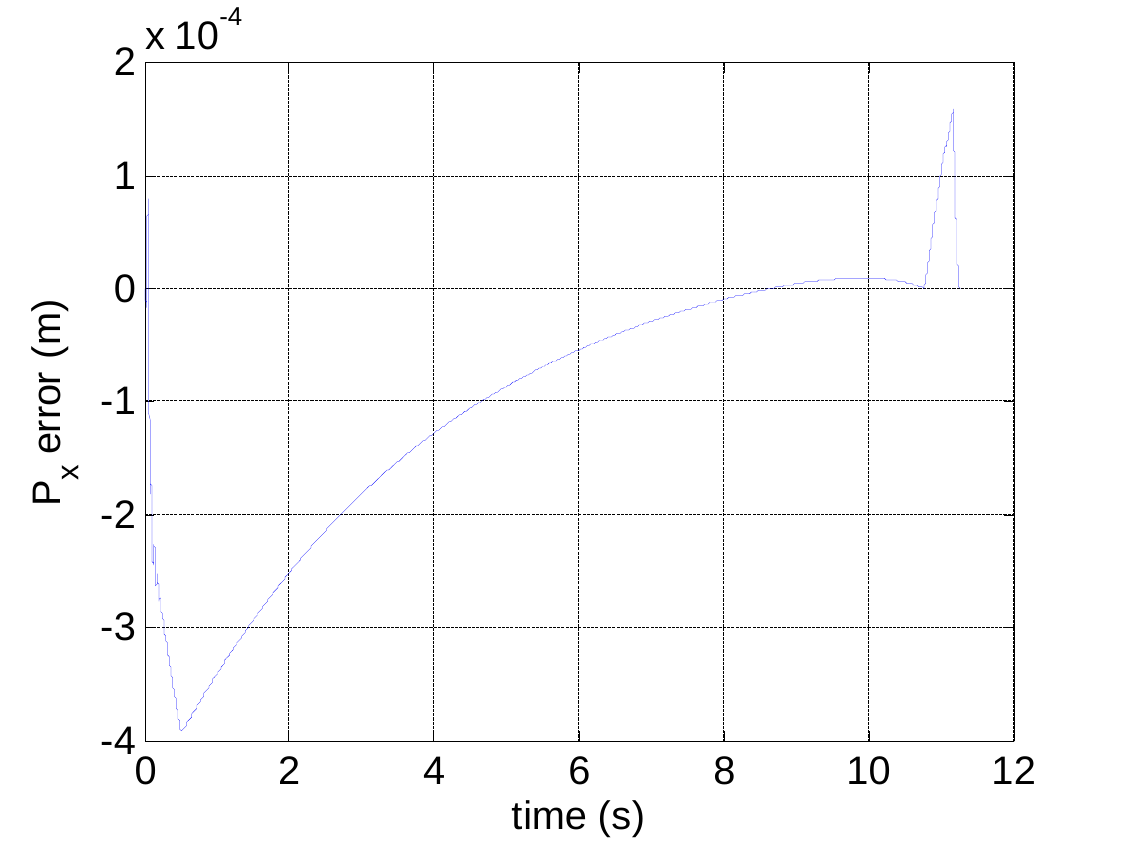}}
\subfloat[]{\includegraphics[width = 5.0cm]{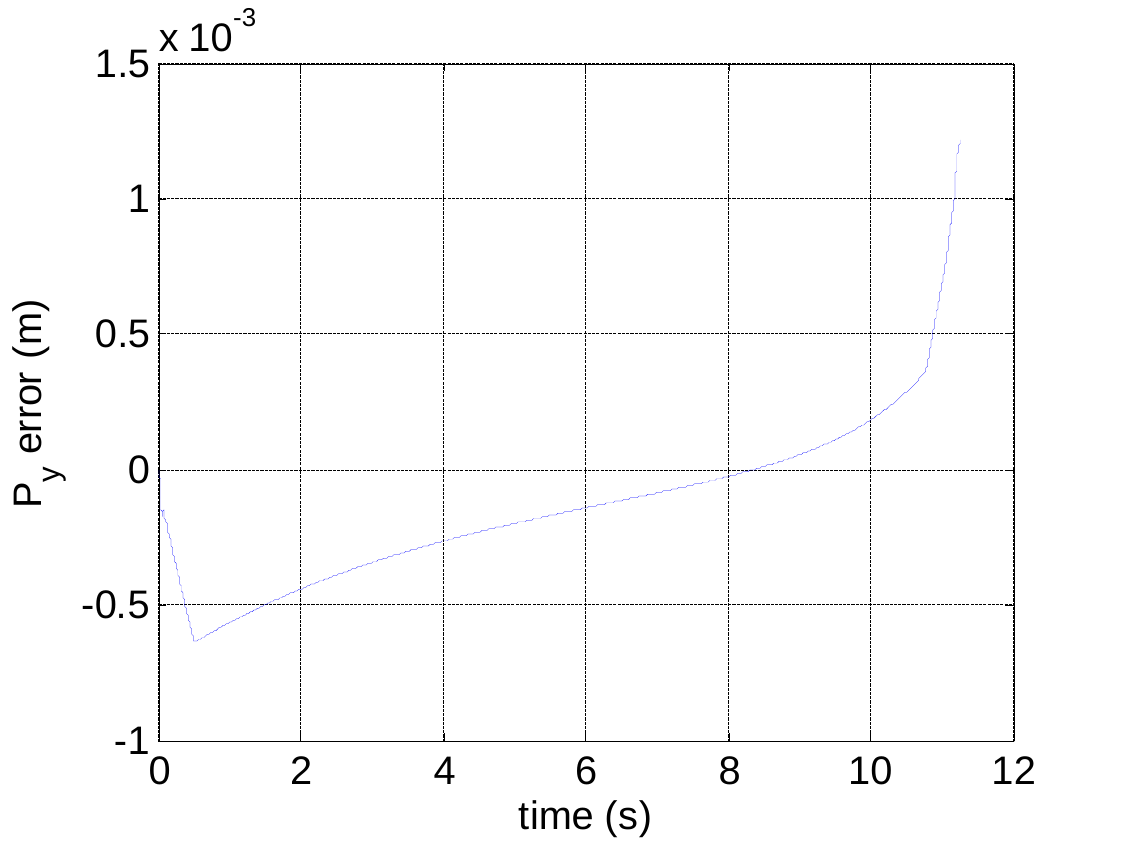}}\\ \vspace{-4mm}
\subfloat[]{\includegraphics[width = 5.0cm]{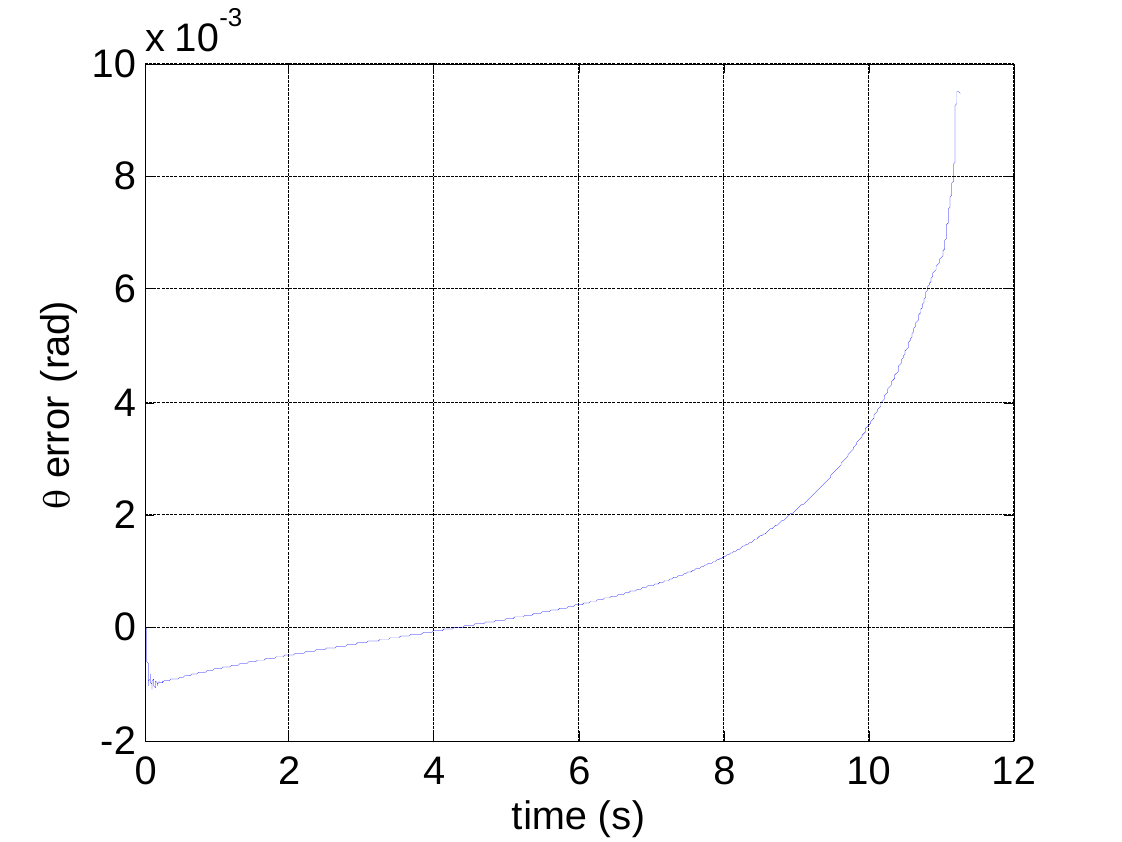}}
\subfloat[]{\includegraphics[width = 5.0cm]{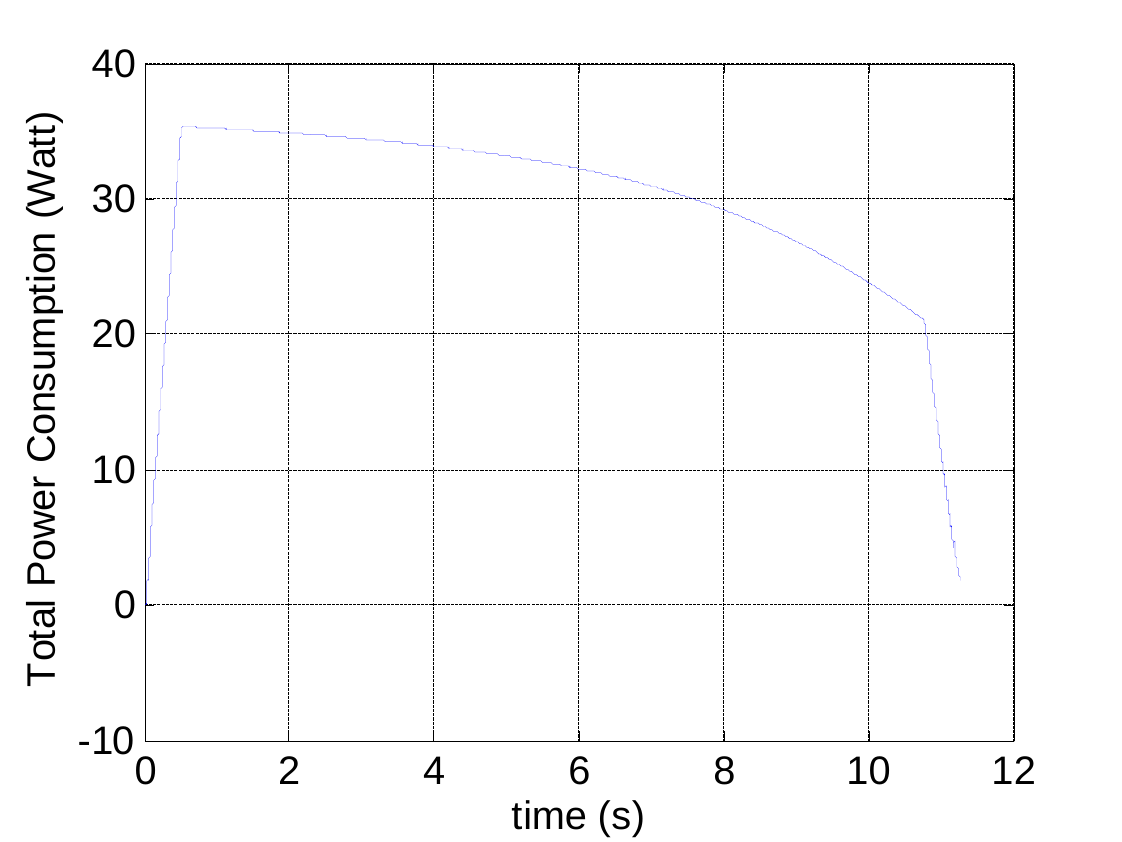}} 
\caption{Simulation results (a) positioning error in cord 1, (b) positioning error in x coordinate, (c) positioning error in y coordinate, (d) angular positioning error, (e) total power consumption of the system}
\label{fig8}
\end{figure}

\begin{table}[h]
\centering
\caption{Parameters of the system.}
\begin{tabular}{|l|l|l|l|}
\hline
\textbf{Parameters}                      & \textbf{Symbols} & \textbf{Values} & \textbf{Unit} \\ \hline
Height of the stand                      & A                & 70              & cm            \\ \hline
Length of the stand                      & B                & 60              & cm            \\ \hline
Height of the movable platform           & a                & 2.8             & cm            \\ \hline
Length of the movable platform           & b                & 15.8            & cm            \\ \hline
Mass of the movable platform             & M                & 0.1             & kg            \\ \hline
Radius of the pulley                     & R                & 2.5             & cm            \\ \hline
Velocity of the movable platform         & $V_p$               & 5               & cm/sec        \\ \hline
Acceleration of the movable platform     & $a_p$               & 10              & cm/sec$^2$       \\ \hline
Inertia of the pulley                    & I                & 3.125$\times10^-5$      & kg$\cdot m^2$          \\ \hline
Inertia of the movable platform          & $I_p$               & 2.617$\times10^-4$      & kg$\cdot m^2$          \\ \hline
Proportional constant used in simulation & $K_p$               & 2000            & constant      \\ \hline
Integral constant used in simulation     & $K_i$               & 500             & constant      \\ \hline
Proportional constant used in experiment & $K_p$               & 0.9             & constant      \\ \hline
Integral constant used in experiment     & $K_i$               & 0.01            & constant      \\ \hline
\end{tabular}
\label{tab1}
\end{table}
Figure~\ref{fig7}a compares actual and reference values of platform’s x and y coordinates and its angle. Results indicate that the actual and reference positions are very close. However, in the angle graph the actual value reaches a little higher than the reference value of zero at the end of simulation. This error is very small (around 0.01 radians), so it is considered acceptable.
Figure~\ref{fig7}b shows the actual and reference trajectories of the center of the platform during the trip. It is seen that x coordinate starts at 0.1 m and ends at 0.3 m, whereas y coordinate starts at 0.1 m and ends at 0.6 m. Furthermore, the actual and reference trajectories closely follow each other.
Figure~\ref{fig8}a-d shows errors between the reference and actual values for cord 1 length, platform coordinates and the platform angle. The maximum error in cord 1 position is about 0.6 mm, which is considered very small. During this trip, a platform positioning accuracy of less than one millimeter is maintained in both x and y directions (see Figures~\ref{fig8}b-\ref{fig8}c). Figure~\ref{fig8}d shows the error in platform angular position, $\theta $. The maximum angular position error is around 0.01 rad, which is barely noticed. Figure~\ref{fig8}e shows total mechanical power requirement to follow the reference path. It is observed that a maximum power of about 35 W is required for this simulated travel path.
\begin{table}[h]
\centering
\caption{RMS errors in simulations (From (10,10) to (30,60) in horizontal position)}
\begin{tabular}{|cl|l|}
\hline
\multicolumn{2}{|c|}{\textbf{Parameters}}                                                         & \textbf{RMS Values} \\ \hline
\multicolumn{1}{|c|}{\multirow{3}{*}{Tracking errors for cord lengths}}            & $\left|L_1-L_{1ref}\right|$     & $2.56\times10^{-4}$ m         \\ \cline{2-3} 
\multicolumn{1}{|c|}{}                                                             & $\left|L_2-L_{2ref}\right|$     & $2.67\times10^{-4}$ m         \\ \cline{2-3} 
\multicolumn{1}{|c|}{}                                                             & $\left|L_3-L_{3ref}\right|$     & $2.81\times10^{-4}$ m         \\ \hline
\multicolumn{1}{|c|}{\multirow{2}{*}{RMS error values for the platform positions}} & $\left|P_x - P_{xref}\right|$ & $1.04\times10^{-4}$ m         \\ \cline{2-3} 
\multicolumn{1}{|c|}{}                                                             & $\left|P_y - P_{yref}\right|$ & $3.06\times10^{-4}$ m         \\ \hline
\multicolumn{2}{|c|}{Orientation error}                                                           & 0.002 rad        \\    \hline
\end{tabular}
\label{tab2}
\end{table}
Table~\ref{tab2} shows the RMS values of positioning errors for the trajectory from (10, 10) to (30, 60) with zero angular position. Among these values, RMS positioning error in y direction has a largest value of about 0.3 mm. Angular positioning error is found as 0.002 rad. These results indicate that the positioning performance of the simulated platform is quite accurate and PI control scheme is sufficient to follow the desired trajectory.
\section{Testing and Experimental Results} In this section, the testing and experimental results will be given and explained. Towards this end, a prototype system is built, tested, and experimental results for two test cases are presented. Motion control is achieved by a PI controller, which acts upon the error between the reference and the actual cord lengths. Proportional and integral constants (Kp and Ki) are listed in Table~\ref{tab1}.
\subsection{Test 1 Level Motion from (50, 10) to (10, 60)} 
\begin{figure}[h!]
\centering
\includegraphics[width=8.0cm]{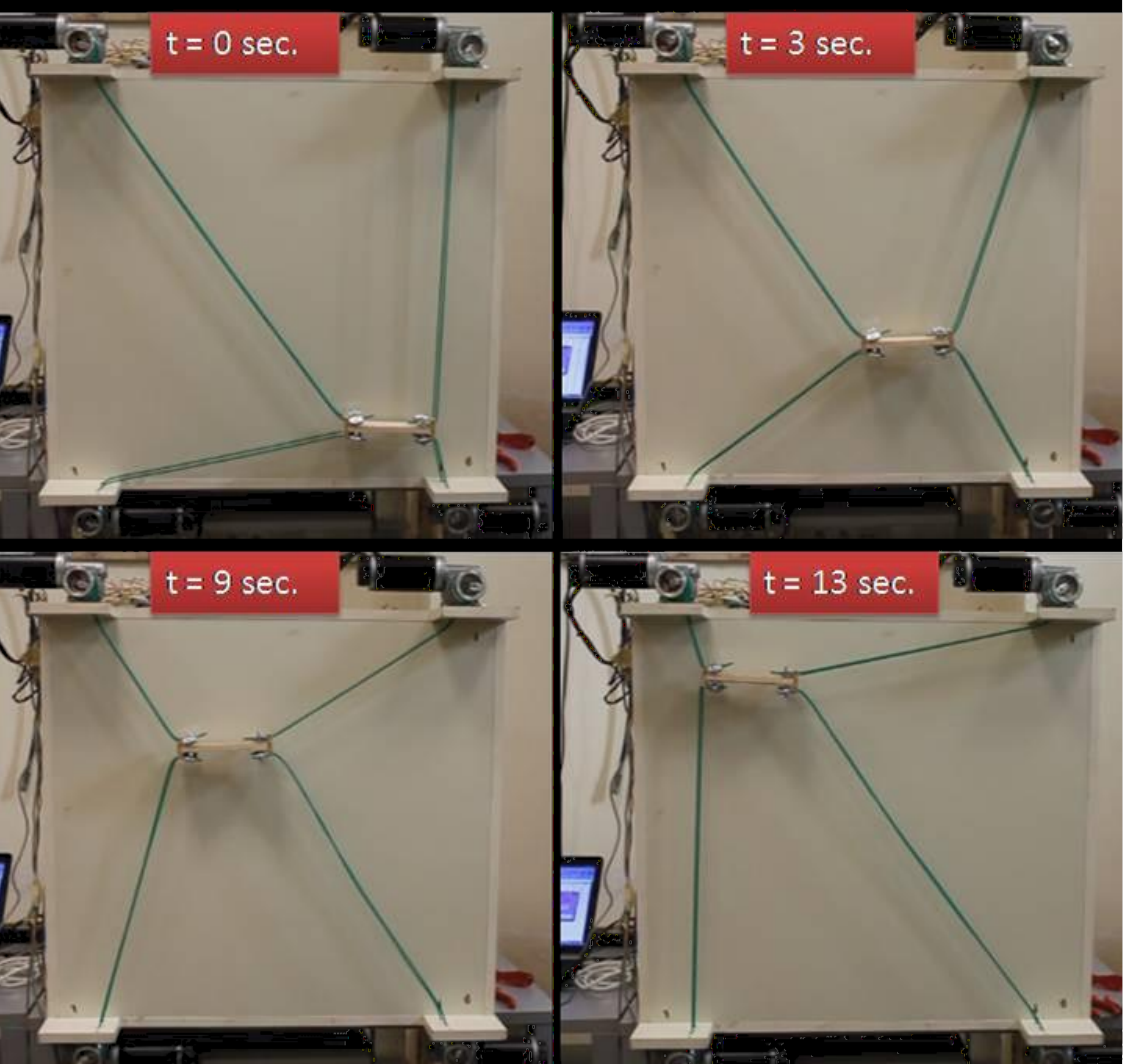}
\caption{Positions of the movable platform in test 1}
\label{fig9}
\end{figure}

\begin{figure}[h!]
\centering
\subfloat[]{\includegraphics[width = 4.5cm]{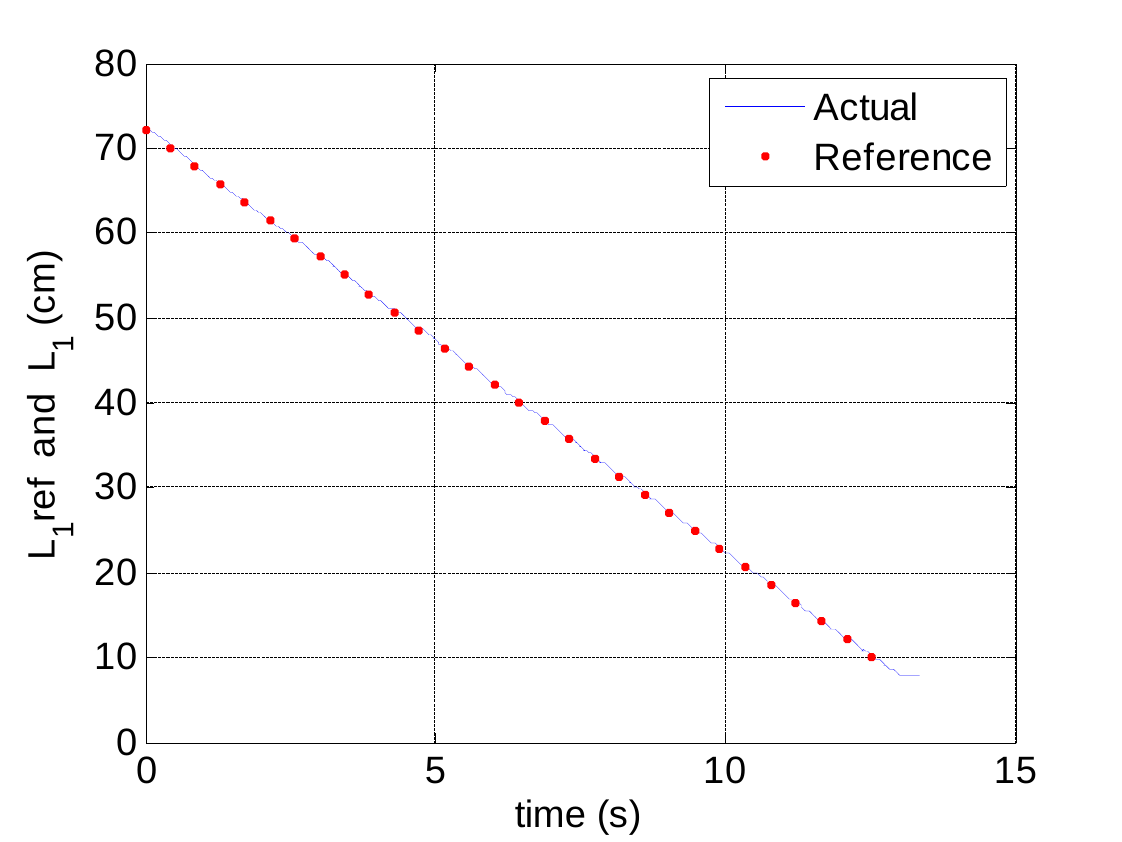}} 
\subfloat[]{\includegraphics[width = 4.5cm]{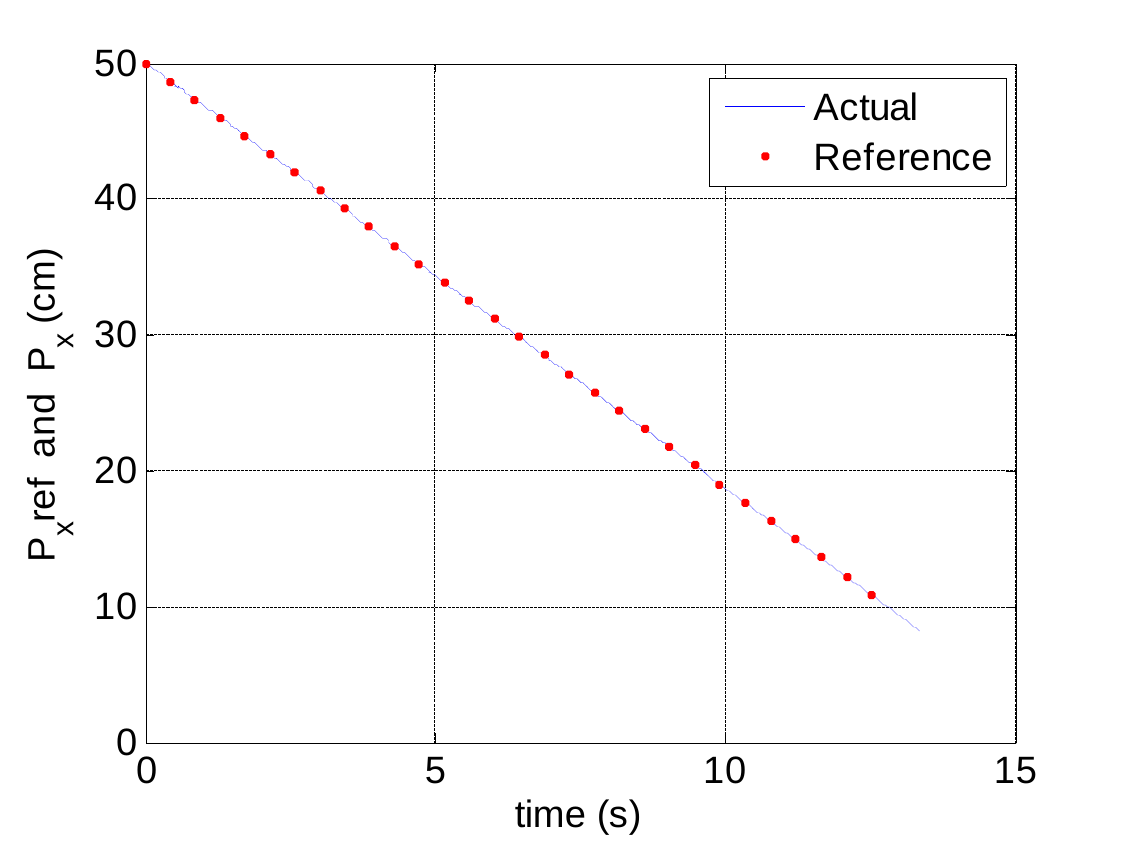}}
\subfloat[]{\includegraphics[width = 4.5cm]{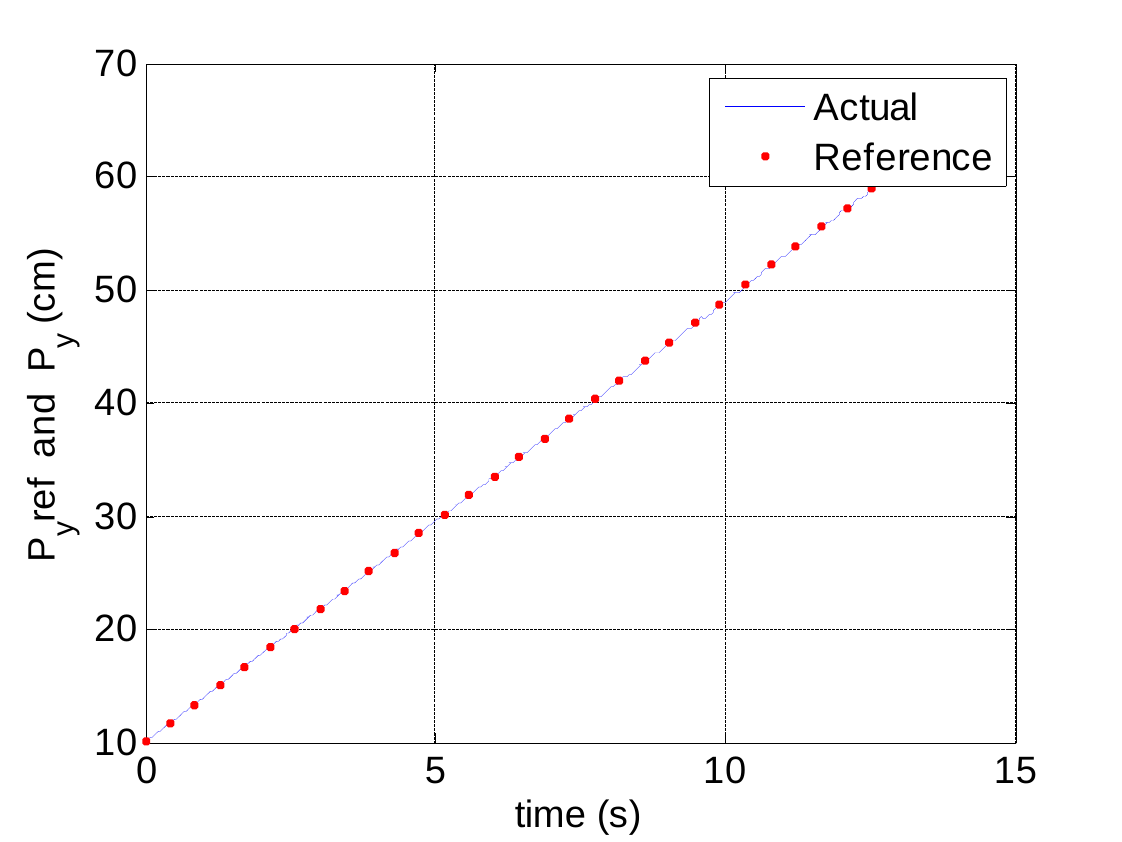}}
\subfloat[]{\includegraphics[width = 4.5cm]{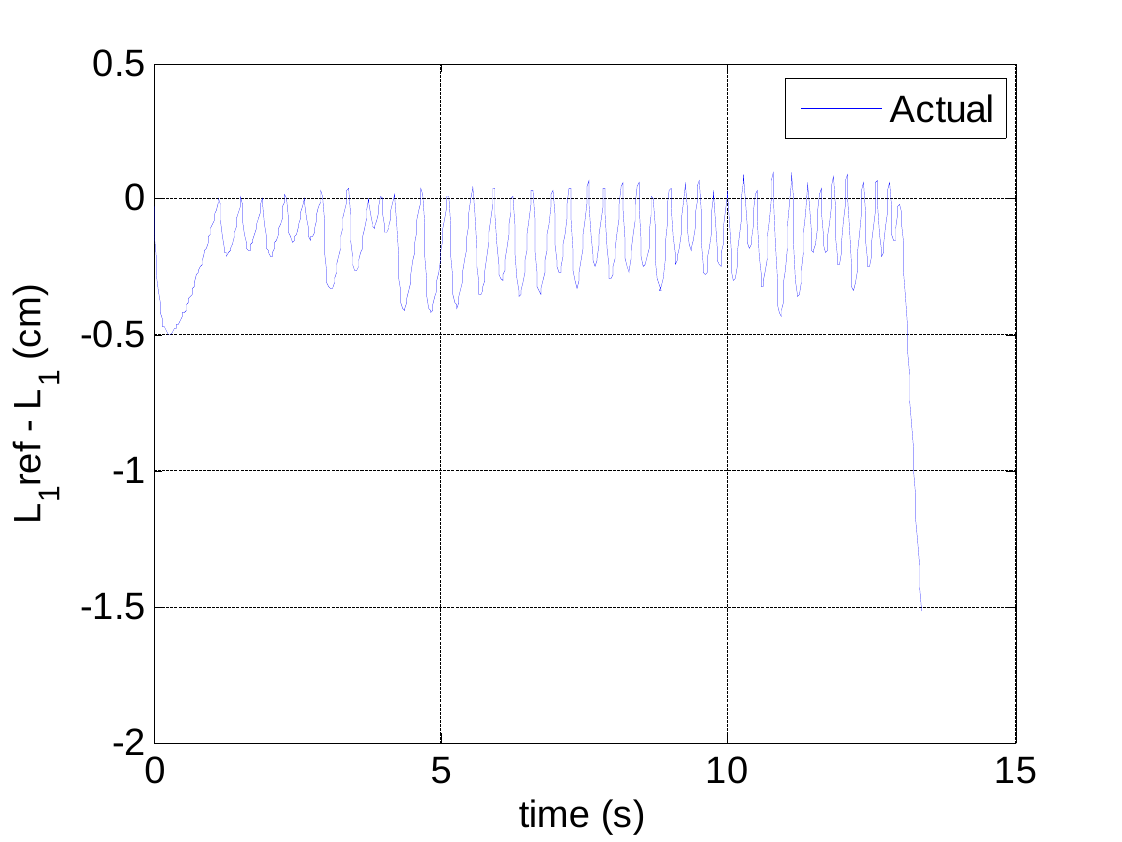}}\\ \vspace{-4mm}
\subfloat[]{\includegraphics[width = 5.0cm]{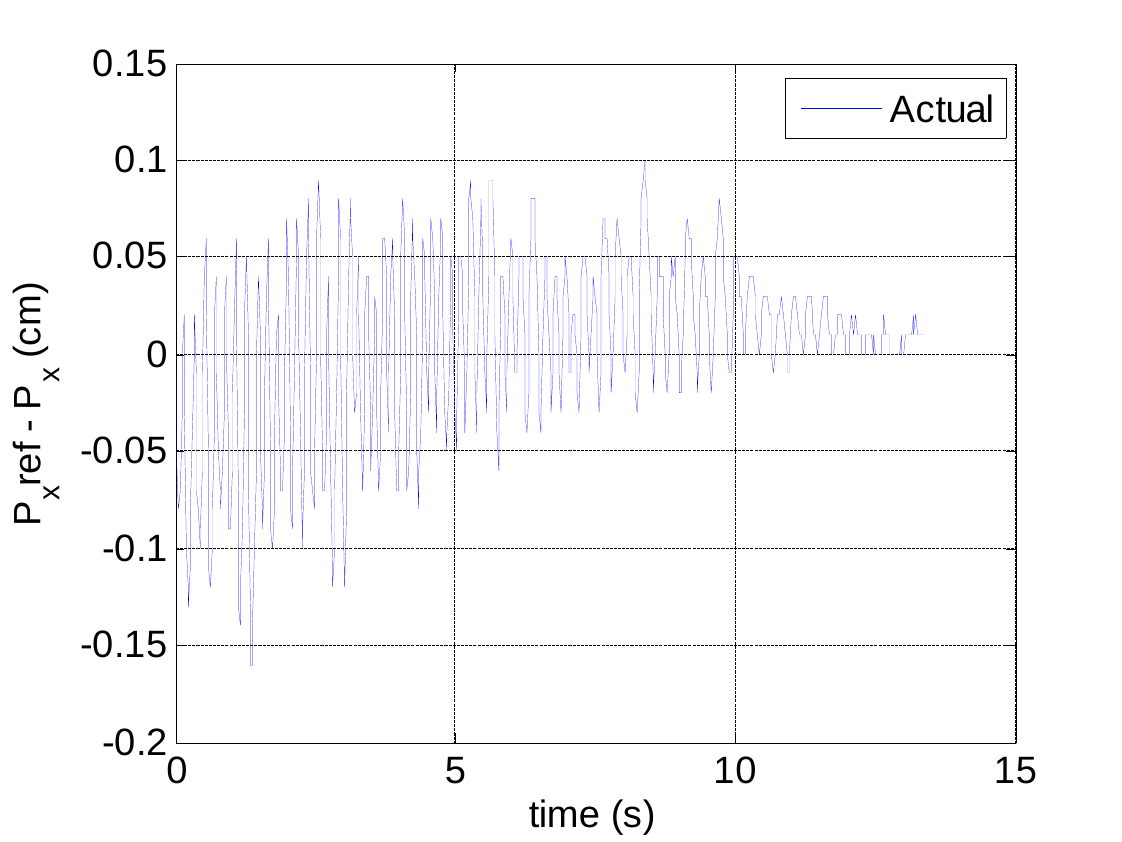}} 
\subfloat[]{\includegraphics[width = 5.0cm]{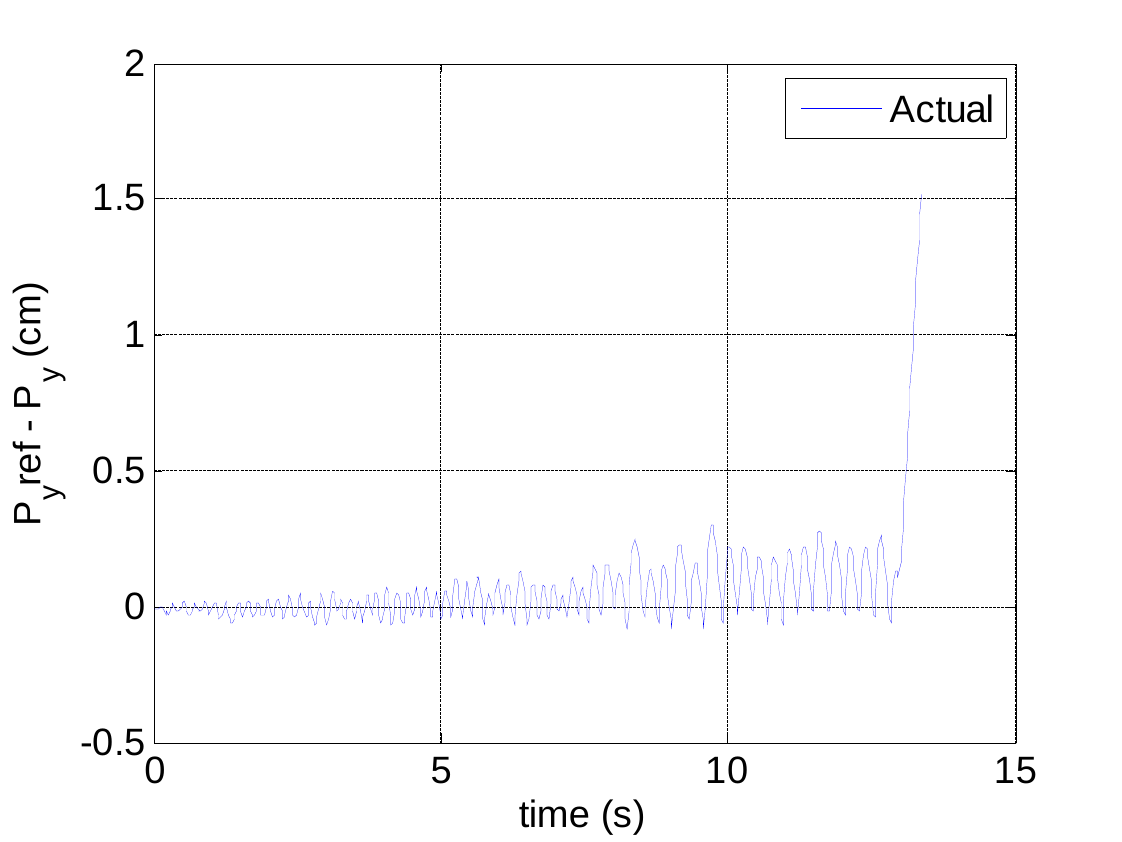}}
\subfloat[]{\includegraphics[width = 5.0cm]{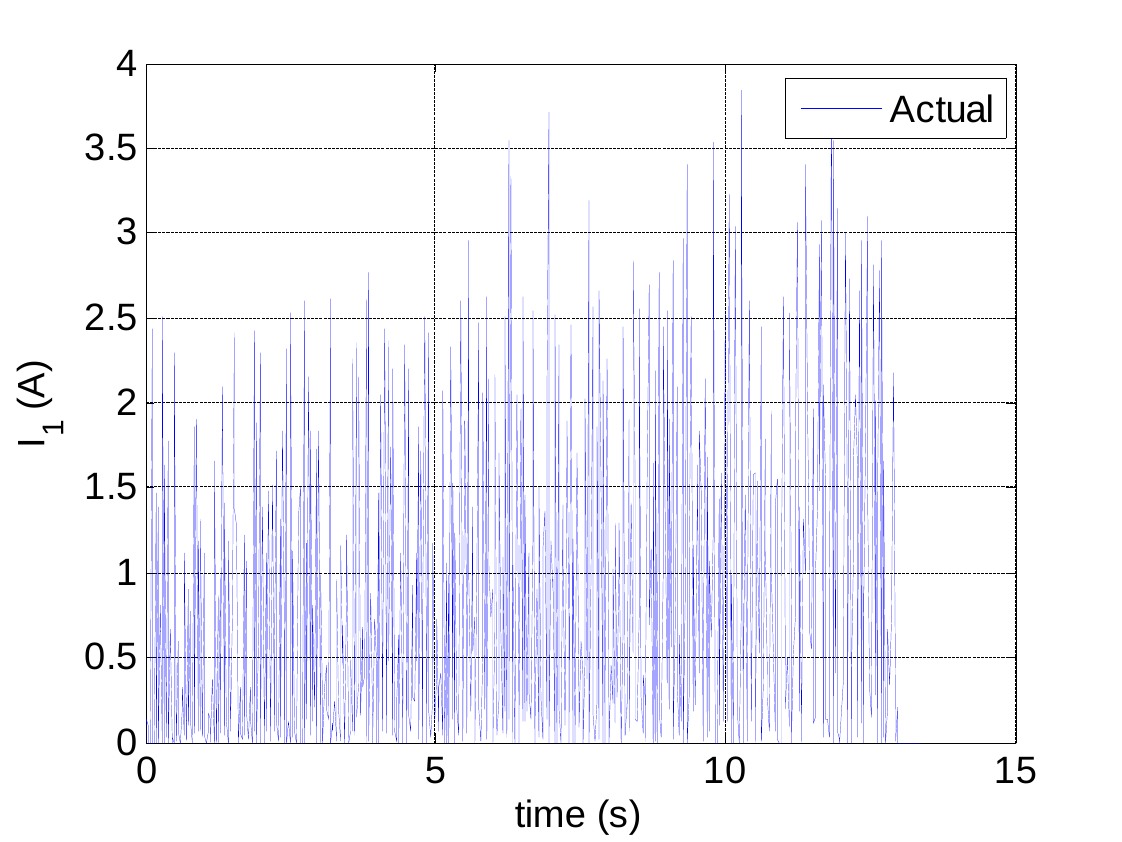}} 
\caption{Performance results in test 1. (a) Actual and reference lengths of cord 1, (b) Actual and reference positions of the platform in x-direction, (c) Actual and reference positions of the platform in y-direction, (d) Error between the actual and reference lengths for cord 1, (e) Error between the actual and reference positions of the platform in x-direction, (f) Error between the actual and reference positions of the platform in y-direction, (g) Current characteristics of motor 1}
\label{fig10}
\end{figure}

An experiment whose initial point (x, y) is (50, 10) cm and final point is (10, 60) cm with no angular motion (${\theta}$ = 0 rad) is run with a platform velocity of 0.05 m/sec and platform acceleration of 0.8 cm/sec2. Views from the motion of the platform are shown in Figure~\ref{fig9}.
Figure~\ref{fig10}a shows reference and actual lengths of cord number 1 in this test. The figure indicates that the actual and reference lengths can closely follow each other. Hence, the control system is able to position the platform accurately.
Figure~\ref{fig10}b and Figure~\ref{fig10}c show actual and reference positions in x and y directions. These graphs also indicate that position of the platform in both x and y directions can be controlled accurately. Figures~\ref{fig10}d, ~\ref{fig10}e, and Figure~\ref{fig10}f show the errors between reference and actual values for cord 1 position, Px, and Py, respectively. The maximum error for cord 1 position is found as 5 mm. In terms of the platform position, the maximum error in x direction is 1.6 mm, whereas in y direction that error is 3 mm. These errors are found to be within reasonable limits. Hence, the positioning accuracy of the prototype is considered quite satisfactory. Figure~\ref{fig10}g shows the current drawn by motor 1. The current signal contains fluctuations, which could be attributed to the nonlinear effects such as the dry friction in motor and gearhead assembly, and the PI controller trying to compensate for these effects. Whenever the control signal changes its sign, the direction of the motor is reversed rapidly and that causes instantaneous jumps in the current signal.
The RMS values of the errors (tracking errors for cord lengths and platform positioning errors) and the currents drawn by the motors are given in Table~\ref{tab3}. Among four, cables 1 and 4 exhibit an error of about 2 mm, whereas cables 2 and 3 exhibit an error of about 1 mm. Those lead to an overall platform positioning error of about 0.5 mm in x direction and 1 mm in y direction, respectively. Current measurements indicate that an average fluctuations of about 1.2 A is present.
\begin{table}[h!]
\centering
\caption{RMS errors in test 1 (From (50,10) to (10,60) in horizontal position)}
\begin{tabular}{|cc|c|}
\hline
\multicolumn{2}{|c|}{\textbf{Parameters}}                                                       & \textbf{RMS Value} \\ \hline
\multicolumn{1}{|c|}{\multirow{4}{*}{Tracking errors for cord lengths}}              & $\left|L_1-L_{1ref}\right|$ & 0.213 cm           \\ \cline{2-3} 
\multicolumn{1}{|c|}{}                                                               & $\left|L_2-L_{2ref}\right|$ & 0.101 cm           \\ \cline{2-3} 
\multicolumn{1}{|c|}{}                                                               & $\left|L_3-L_{3ref}\right|$ & 0.113 cm           \\ \cline{2-3} 
\multicolumn{1}{|c|}{}                                                               & $\left|L_4-L_{4ref}\right|$ & 0.204 cm           \\ \hline
\multicolumn{1}{|c|}{\multirow{2}{*}{RMS   error values for the platform positions}} & $\left|P_x-P_{xref}\right|$ & 0.047 cm           \\ \cline{2-3} 
\multicolumn{1}{|c|}{}                                                               & $\left|P_y-P_{yref}\right|$ & 0.103 cm           \\ \hline
\multicolumn{1}{|c|}{\multirow{4}{*}{Average   fluctuations in currents}}            & $I_1$ & 1.254 A            \\ \cline{2-3} 
\multicolumn{1}{|c|}{}                                                               & $I_2$ & 1.218 A            \\ \cline{2-3} 
\multicolumn{1}{|c|}{}                                                               & $I_3$ & 1.155 A            \\ \cline{2-3} 
\multicolumn{1}{|c|}{}                                                               & $I_4$ & 1.226 A            \\ \hline
\end{tabular}
\label{tab3}
\end{table}
\subsection{Test 2 Angular Motion from (10, 30) to (50, 30)} 
\begin{figure}[h!]
\centering
\includegraphics[width=8.0cm]{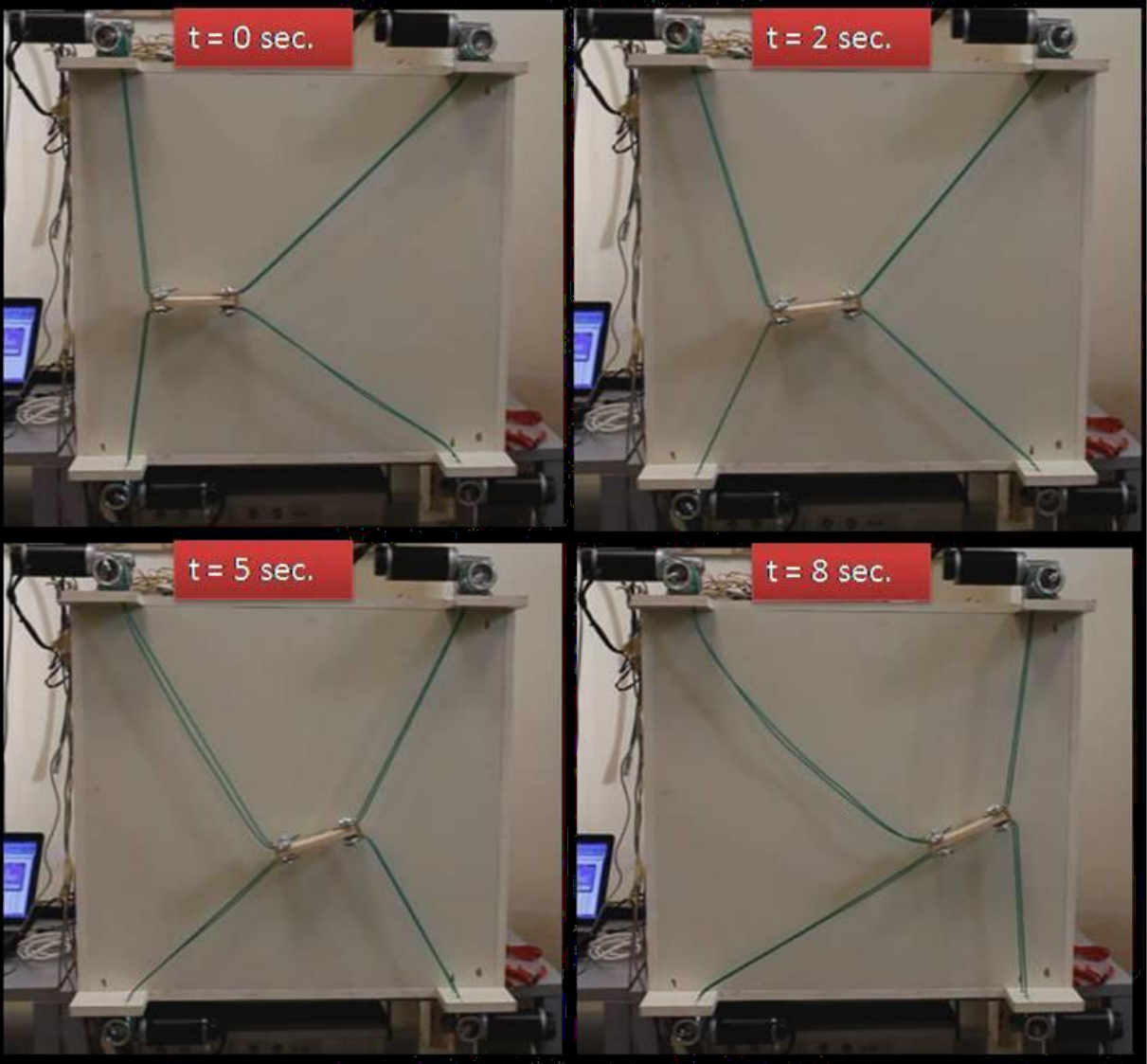}
\caption{Positions of the movable platform in test 2}
\label{fig11}
\end{figure}

\begin{figure}[h!]
\centering
\subfloat[]{\includegraphics[width = 6.0cm]{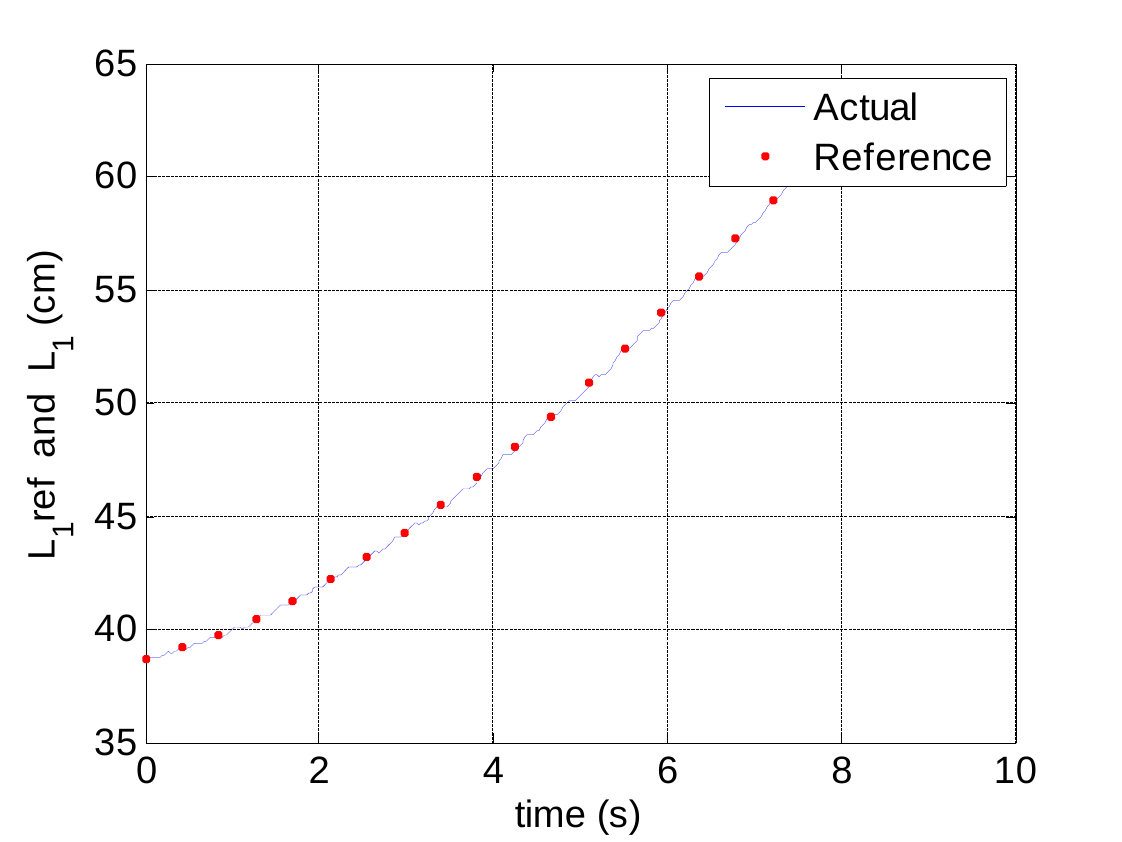}} 
\subfloat[]{\includegraphics[width = 6.0cm]{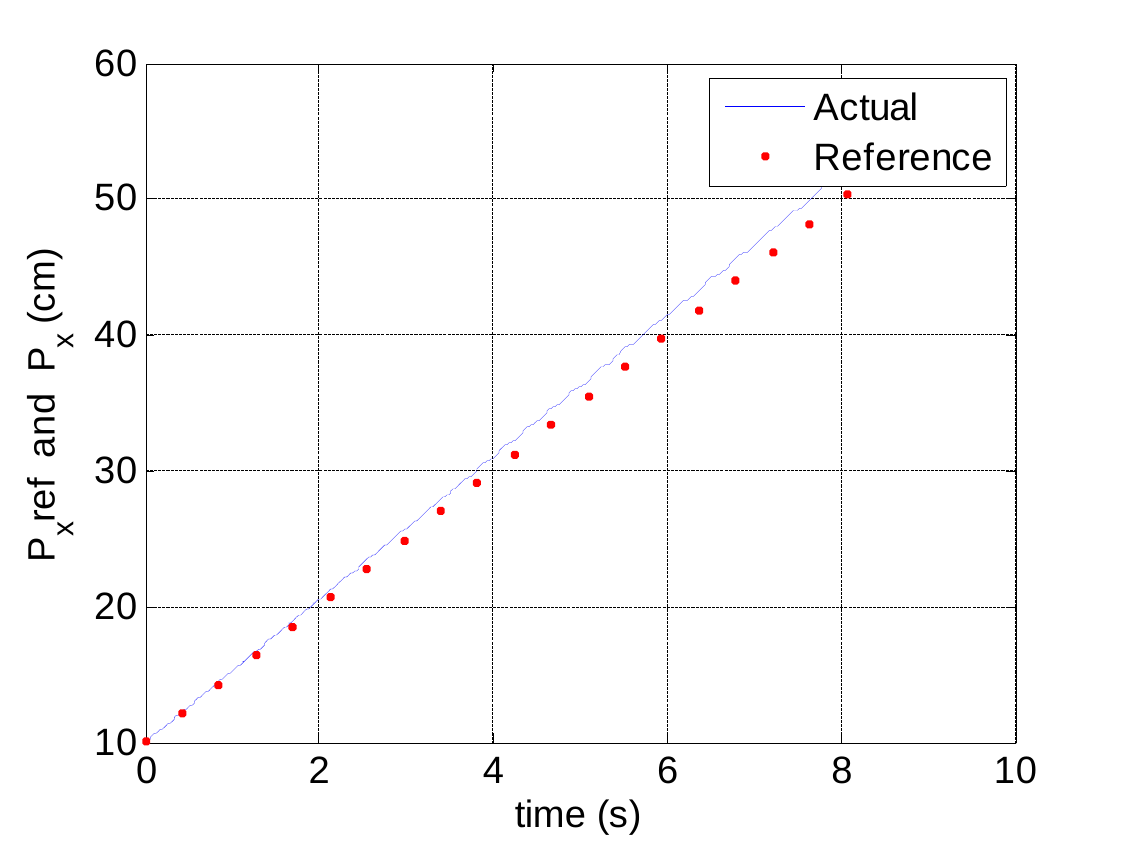}}
\subfloat[]{\includegraphics[width = 6.0cm]{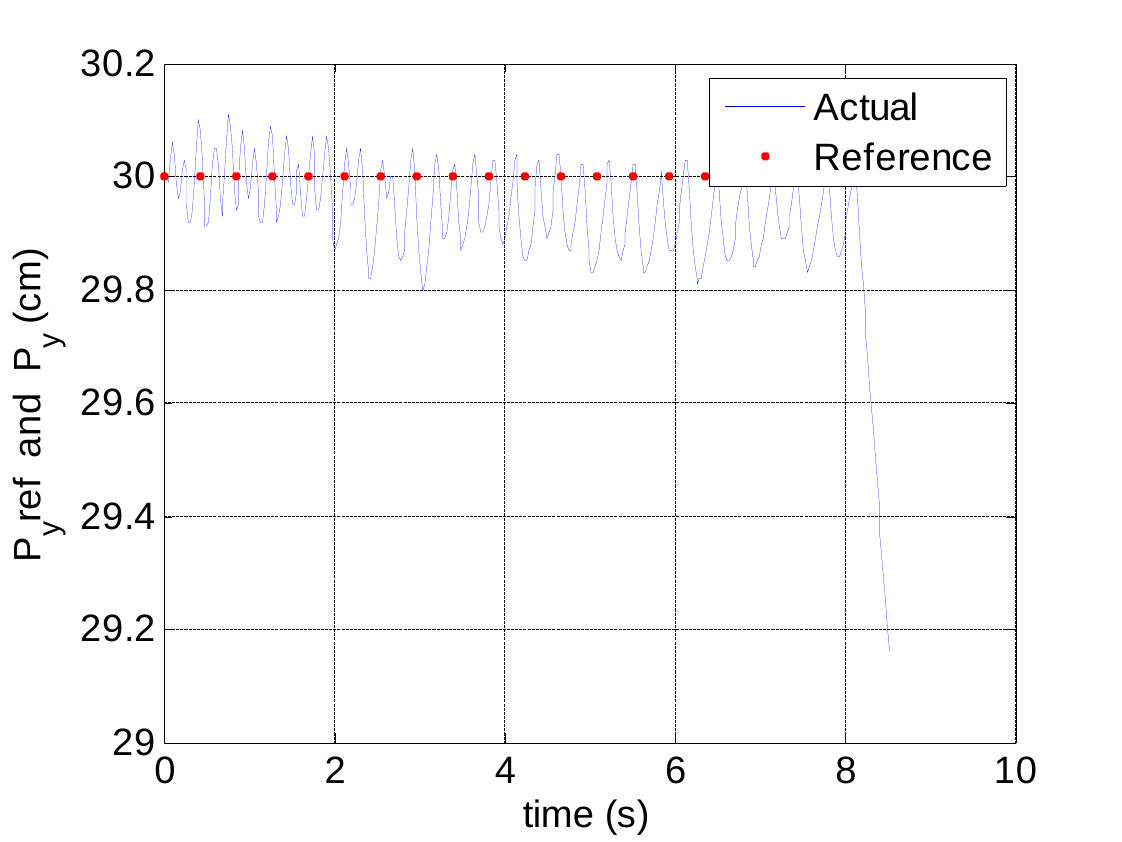}}\\
\subfloat[]{\includegraphics[width = 6.0cm]{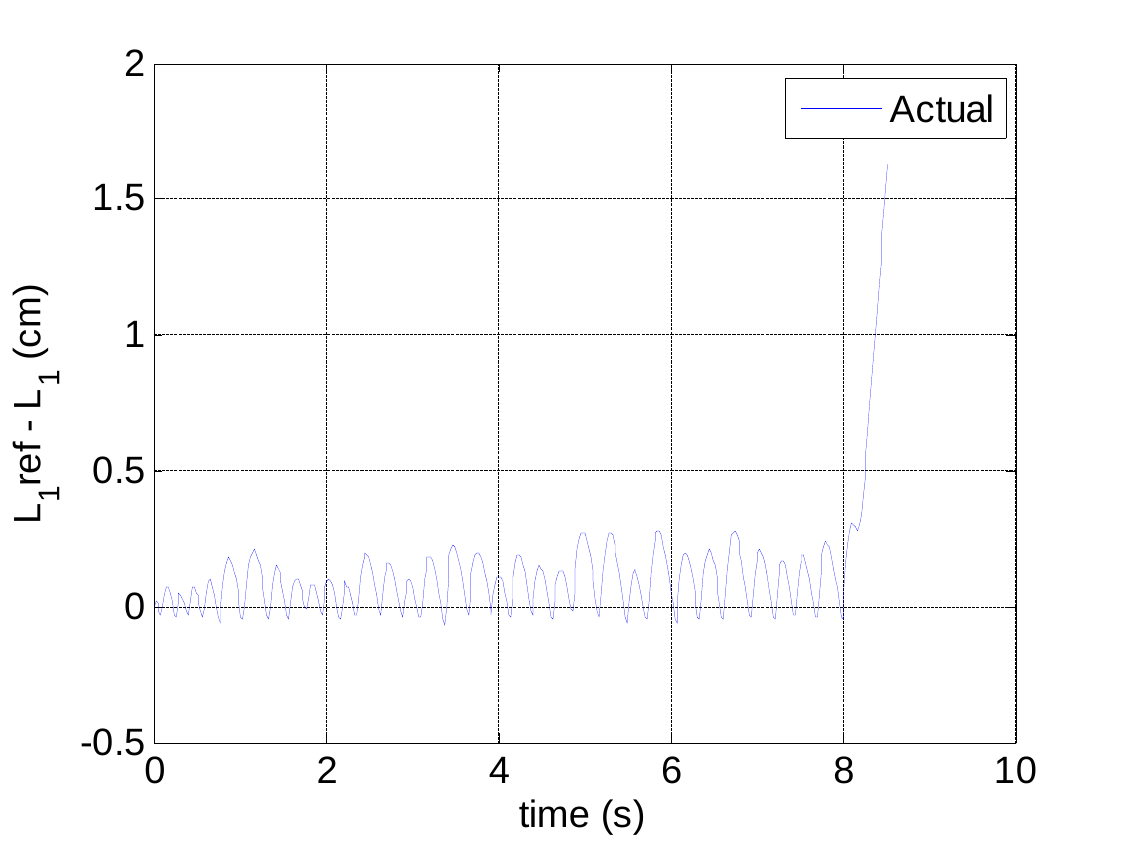}}
\subfloat[]{\includegraphics[width = 6.0cm]{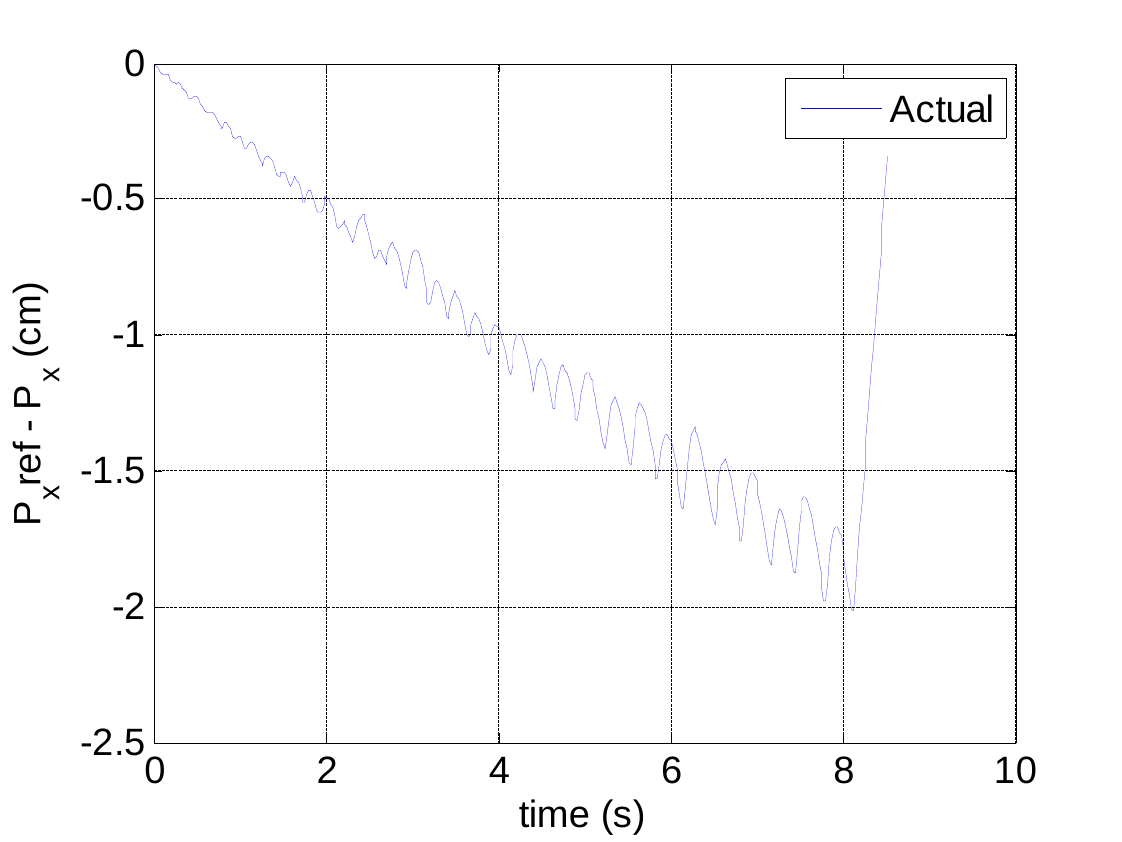}} 
\subfloat[]{\includegraphics[width = 6.0cm]{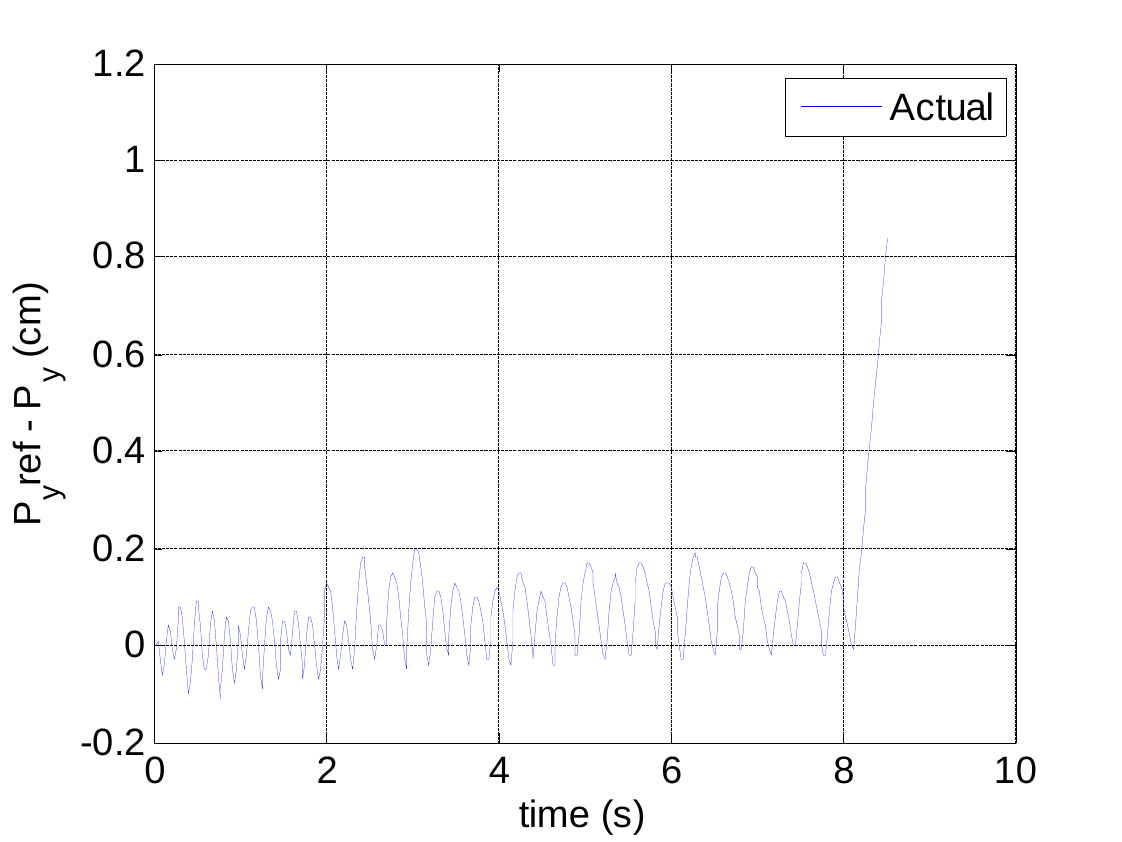}}\\
\subfloat[]{\includegraphics[width = 6.0cm]{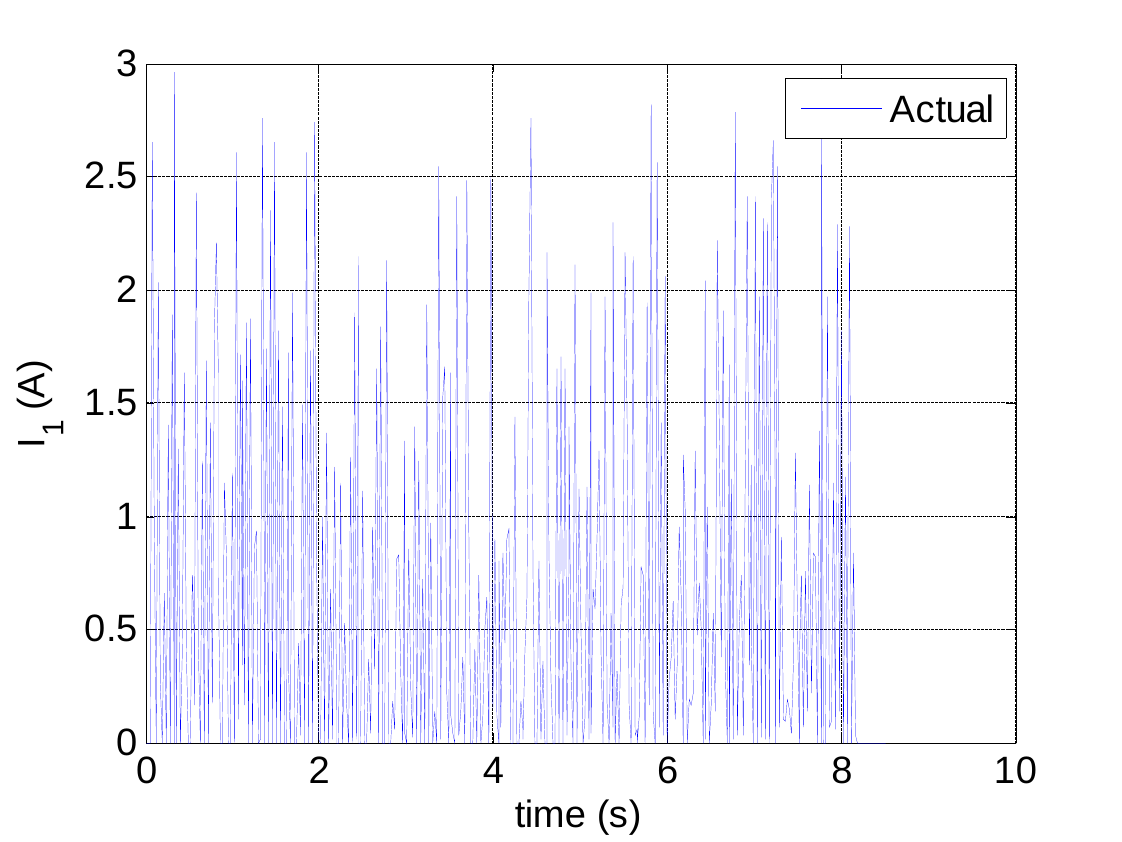}} 
\caption{Performance results in test 2. (a) Actual and reference lengths of cord 1, (b) Actual and reference positions of the platform in x-direction, (c) Actual and reference positions of the platform in y-direction, (d) Error between the actual and reference lengths for cord 1, (e) Error between the actual and reference positions of the platform in x-direction, (f) Error between the actual and reference positions of the platform in y-direction, (g) Current characteristics of motor 1}
\label{fig12}
\end{figure}

The following results are obtained for the case when the movable platform starts at a level position and reaches an angular position of 45 degrees. Pictures taken while the platform is moved from (10, 30) cm to (50, 30) cm, with an angle command of 0 to 45 degrees are given in Figure~\ref{fig11}. 
Results from this test are shown in Figure~\ref{fig12}. Reference tracking performance for cord 1 positions is found quite satisfactory (Figure~\ref{fig12}a). The maximum tracking error in cord 1 position is 2.8 mm (Figure~\ref{fig12}d). Platform positioning performance in y direction is acceptable (Figure~\ref{fig12}c), but its x direction performance is poor (Figure~\ref{fig12}b). The positioning error in x direction builds up during the test and reaches at a maximum value of 20.2 mm (Figure~\ref{fig12}e). The positioning error in y direction remains within a maximum of 2 mm, which is considered reasonable (Figure~\ref{fig12}f). Current drawn by motor 1 during test 2 is shown in Figure~\ref{fig12}g, which exhibits oscillatory behavior.
The RMS values for the errors obtained in test 2 are shown in Table~\ref{tab4}. The positioning error for the cords remains below an RMS value of 1.5 mm. Platform positioning errors are about 0.9 mm (RMS) in y direction and 11.4 mm (RMS) in x direction. Current consumption of the motors is below 1.2 A (RMS). These results demonstrate that performance of prototype movable platform is satisfactory.
\begin{table}[h!]
\centering
\caption{RMS errors in test 2 (From (10,30) to (50,30) with 0-45 degree angle command)}
\begin{tabular}{|c|cc|}
\hline
\textbf{Parameters}                                          & \multicolumn{2}{c|}{\textbf{RMS Value}}  \\ \hline
\multirow{4}{*}{RMS tracking errors for the 4 cords}         & \multicolumn{1}{c|}{$\left|L_1-L_{1ref}\right|$} & 0.132 cm \\ \cline{2-3} 
                                                             & \multicolumn{1}{c|}{$\left|L_2-L_{2ref}\right|$} & 0.142 cm \\ \cline{2-3} 
                                                             & \multicolumn{1}{c|}{$\left|L_3-L_{3ref}\right|$} & 0.147 cm \\ \cline{2-3} 
                                                             & \multicolumn{1}{c|}{$\left|L_4-L_{4ref}\right|$} & 0.123 cm \\ \hline
\multirow{2}{*}{RMS error values for the platform positions} & \multicolumn{1}{c|}{$\left|P_x-P_{xref}\right|$} & 1.141 cm \\ \cline{2-3} 
                                                             & \multicolumn{1}{c|}{$\left|P_y-P_{yref}\right|$} & 0.090 cm \\ \hline
\multirow{4}{*}{Average fluctuations in currents}            & \multicolumn{1}{c|}{$I_1$}       & 1.058 A  \\ \cline{2-3} 
                                                             & \multicolumn{1}{c|}{$I_2$}       & 1.228 A  \\ \cline{2-3} 
                                                             & \multicolumn{1}{c|}{$I_3$}       & 1.181 A  \\ \cline{2-3} 
                                                             & \multicolumn{1}{c|}{$I_4$}       & 1.002 A  \\ \hline
\end{tabular}
\label{tab4}
\end{table}
\subsection{Discussion and Contributions} Presented results of the motion control experiments indicate feasibility of the proposed 3-DOF cable-suspended scaffolding system to operate on buildings and structures. Particularly, capability of operating on the vertical plane is one distinguishing feature of this system. Laboratory-scale prototype can be built on a fairly low budget. Future studies can be directed towards building a real-size system and application areas that require precise positioning on the vertical plane such as cleaning and painting operations.
\section{Conclusions} This paper develops a solution to the scaffolding problem for buildings and large structures. Towards this purpose, a novel design has been proposed, and its behavior has been validated by simulations and by building a laboratory-scale prototype. The proposed system consists of four motors and a movable platform, which is driven by cords attached to these motors. Said system offers advantages over the current scaffolding systems such as ease of assembly, compactness, and high dexterity. Furthermore, this system can be used for other purposes such as surface cleaning, painting on the vertical planes, plastering and performing pick-and-place operations. The proposed solution is capable of moving not only along straight line paths but also along curved ones. In this paper, a prototype movable scaffolding system, which can travel along any commanded direction, has been tested successfully. Results obtained from both simulations and actual tests confirmed that the proposed mechatronic system has quite accurate positioning performance.

\section*{Acknowledgments} Authors are thankful to Ömer Sönmez and Cem Gürdal for their contributions in building the test setup and conducting the experiments.

\end{document}